\documentclass[aps,prx,twocolumn,10pt]{revtex4-1}

\usepackage{amssymb}
\usepackage{amsbsy}
\usepackage{amsmath}
\usepackage{epsfig}
\usepackage{graphicx}
\usepackage{graphics}
\usepackage{array}
\usepackage{color}
\usepackage{textcomp}
\usepackage{cleveref}
\usepackage[body={18.5cm,24cm},top={2.cm},left={1.7cm}]{geometry}

\let\a=\alpha \let\b=\beta \let\g=\gamma \let\d=\partial 
   
\let\l=\lambda   
\let\s=\sigma    

\let\w=\omega  \let\D=\Delta

\def\nn{\nonumber}

\def\bpm{\begin{pmatrix}}
\def\epm{\end{pmatrix}}
\def\be{\begin{equation}}
\def\ee{\end{equation}}
\def\bea{\begin{eqnarray}}
\def\eea{\end{eqnarray}}
\def\ba{\begin{array}}
\def\ea{\end{array}}
\def\del{\delta}

\def\td{\tilde}

\def\ep{{\epsilon}}

\def\bfr{\mathbf{r}}

\newcommand{\bp}{{\bf p}}
\newcommand{\bq}{{\bf q}}

\newcommand{\trasp}{\mathsf{T}}

\newcommand{\mathsym}[1]{{}}
\newcommand{\unicode}[1]{{}}

\newcommand{\txa}{\mathrm{a}}
\newcommand{\txb}{\mathrm{b}}

\newcommand{\divv}{\mathop{\rm div}\nolimits}

\begin{document}
\title{Negative viscosity and eddy flow of imbalanced electron-hole liquid in graphene}
\author{Hong-Yi Xie} 
\affiliation{Department of Physics, University of Wisconsin-Madison, Madison, Wisconsin 53706, USA}
\author{Alex Levchenko}
\affiliation{Department of Physics, University of Wisconsin-Madison, Madison, Wisconsin 53706, USA}
\date{December 28, 2018}
\begin{abstract}
We present a hydrodynamic theory for electron-hole magnetotransport in graphene incorporating carrier-population imbalance, energy, and momentum relaxation processes. We focus on the electric response and find that the carrier and energy imbalance relaxation processes strongly modify the shear viscosity, so that an effective viscosity can be negative in the vicinity of charge neutrality. We predict an emergent eddy flow pattern of swirling currents and explore its manifestation in nonlocal resistivity oscillations in a strip of graphene driven by a source current.
\end{abstract}
\maketitle


\section{Introduction} 

In hydrodynamics an \textit{eddy phenomenon} is a particular large-scale turbulentlike motion of the fluid with a distinct swirling pattern of the flow velocity. It is often discussed in conjunction with the concept of negative viscosity, synonymously called eddy or turbulent viscosity, that has roots going back to the early studies of Reynolds [\onlinecite{Reynolds}]. In contrast to kinematic shear viscosity, which describes the physical properties of the fluid, the eddy viscosity describes the properties of the flow itself. For that reason it is sign indefinite and could be negative, unlike the shear viscosity, which is strictly positive, as dictated by the second law of thermodynamics for irreversible processes [\onlinecite{LLbook}]. The negative viscosity effects are counterintuitive. The classical text of Starr [\onlinecite{Starr}] contains a number of essays summarizing some empirical facts and describes the spectacular manifestations of eddies in the geophysical context of Earth's atmosphere and oceanic streams, the Sun's photosphere, and spiraling galaxies. In analytical models, perhaps the simplest hydrodynamic system exhibiting a negative viscosity effect is the so-called Kolmogorov flow: the two-dimensional flow of a viscous liquid induced by a unidirectional external force field periodic in one of the coordinates [\onlinecite{Sivashinsky}]. The stability of such flows has been extensively investigated by taking into account higher-order gradient and nonlinear terms in Navier-Stokes equations, and by the direct numerical modeling [\onlinecite{Chaves}]. An emergent regime of negative viscosity was also found in magnetohydrodynamics of tokamak plasmas [\onlinecite{Biskamp}], ferrofluids [\onlinecite{Bacri}], and in the description of Rossby wave turbulence [\onlinecite{Chechkin}]. 

Is it possible to have an analog of these effects in strongly correlated electron systems? The idea that electrons in solids can flow hydrodynamically was put on firm footing by Gurzhi [\onlinecite{Gurzhi}]. It took, however, several decades for the manifestations of electronic viscous effects to be observed in macroscopic transport experiments [\onlinecite{Molenkamp-1},\onlinecite{Molenkamp-2}]. The reason has to do with the fact that, typically, low-temperature transport in the usual materials is dominated by disorder, which is incompatible with the hydrodynamic picture as electron-impurity scattering quickly relaxes momentum. Raising the temperature leads to a shorter electron-electron scattering time and thus sufficiently fast equilibration of the electron liquid; however, at elevated temperatures electron-phonon scattering begins to dominate, leading to both momentum and energy relaxations. As a consequence, the hydrodynamic regime can be expected only in an intermediate range of temperatures in very clean samples where electronic equilibration occurs on length scales that are short compared to those of momentum and energy relaxations. This is, in practice, difficult to realize in most materials. A wealth of transport data extracted from measurements on two-dimensional electron systems in high- mobility semiconductor devices with low electron densities is presented in the review in [\onlinecite{Spivak}], where arguments were put forward that multiple observed features can be understood by invoking hydrodynamic effects. Recently, various signatures of hydrodynamic flow, such as current whirlpools and anomalous thermal conductivity and thermopower, were observed and explained in monolayer graphene [\onlinecite{Crossno,Bandurin,Ghahari,Kumar,Levitov,Polini}] and palladium cobaltate [\onlinecite{Mackenzie}].  Monolayer graphene (MLG) on hexagonal boron nitride (hBN) represents essentially a unique system in which, due to its purity, electrons can be brought into the hydrodynamic regime over a fairly wide range of temperatures, from $50 \, \mathrm{K}$ to practically room temperature, and furthermore, the electron-electron scattering length can be controlled by tuning the carrier concentration using a gate electrode. In this work we report on the possibility of an eddy pattern formation in an imbalanced electron-hole liquid in graphene and develop a corresponding microscopic theory. Special attention is paid to determining the region in the density-temperature-field phase diagram where this effect is strongest. We discuss experimentally relevant geometry and give concrete predictions for the manifestations of Dirac fluid eddies in the nonlocal magnetotransport measurements.     

The rest of the paper is organized as follows. In Sec. \ref{Sec-Hydro} we formulate the generic hydrodynamic transport theory applicable to the electron-hole liquid in MLG subject to external magnetic field. Additional details for this section are provided in Appendix \ref{App-Hydro}. We analyze linearized hydrodynamic equations in Sec. \ref{Sec-Linear-Hydro} with the microscopic coefficients and relaxation rates computed from the underlying kinetic theory, which is sketched in Appendix \ref{App-Relaxation}. In Sec. \ref{Sec-Stream} we derive the stream function equation for the hydrodynamic flow and solve it with a Fourier transform in Sec. \ref{Sec-Eddy} to reveal the regime of the eddy flow. In Appendix \ref{App-Fourier} we present the same computation carried out for different boundary conditions. We summarize our findings in Sec. \ref{Sec-Fin} with an angle on recently published related work and perspectives for future studies.    
 
\section{Hydrodynamic equations}\label{Sec-Hydro}

Assuming fast equilibration of electron-hole plasma due to strong e-e(h) inelastic Coulomb collisions, we express the carrier current densities and energy-momentum tensor in terms of local thermodynamic functions, hydrodynamic velocity, and dissipative deviations from local equilibrium. The resulting hydrodynamic equations [\onlinecite{Mueller2008,Foster2009,Naro2015,Lucas2018}]
\begin{subequations} \label{hyd-eqs}
\begin{align}
& \d_t \mathsf{\rho} + \divv\mathbf{J}= 0, \quad \d_t n+ \divv\mathbf{P} = \mathcal{I},  \label{cn-cons} \\
& \left(h/v_F^2\right) \left( \d_t + \mathbf{u}\cdot \nabla \right) \mathbf{u} = - \nabla \mathcal{P}+ \boldsymbol{l} + \boldsymbol{f} - \nabla \cdot \hat{\theta}, \label{ns-eq} \\
& T \left[\d_t s + \divv\left( s \, \mathbf{u} - \frac{\nu}{T} \mathbf{p} + \frac{\mu}{e T} \mathbf{j}  \right) \right] =  \varpi - \mathbf{u} \cdot \boldsymbol{f} - \nu \mathcal{I} \nn \\
& + \frac{1}{e}\left[\boldsymbol{L} + T \nabla \left( \frac{\mu}{T} \right) \right] \cdot \mathbf{j} - T \nabla \left( \frac{\nu}{T} \right) \cdot \mathbf{p} - \hat{\theta} : \nabla \mathbf{u}, \label{en-prod}
\end{align} 
\end{subequations}
include the charge and carrier continuity, Navier-Stokes, and entropy production equations (throughout the paper we use units of $\hbar=k_B=1$). Here $v_F$ is the Fermi velocity, $e>0$ is the elementary charge, and $\rho$, $n$, $h$, and $s$ are the proper charge, carrier number, enthalpy, and entropy density, respectively. Finally, $\mu$, $\nu$, and $T$ are, respectively, the relative chemical potential, imbalance chemical potential, and temperature. The total pressure $\mathcal{P}$ satisfies the equation of state $\mathcal{P} = h/3$, which is a consequence of relativistic scale invariance. 
In Eq.~\eqref{hyd-eqs}, the charge current density $\mathbf{J}$ and carrier current density $\mathbf{P}$ are parametrized as $\mathbf{J} = \rho \mathbf{u}  + \mathbf{j}$, and  $\mathbf{P} = n \mathbf{u}  + \mathbf{p}$, where $\mathbf{u}$ is the fluid velocity, $\mathbf{j}$ and $\mathbf{p}$ are the dissipative fluctuations, and $\hat{\theta}$ describes the dissipative part of the stress tensor manifesting the viscous effects. We have assumed the limit $\mathbf{u}^2/v_F^2 \ll 1$ and ensured that the proper densities receive no dissipative corrections, so that the dissipative fluctuations are orthogonal to the fluid velocity (see Appendix \ref{App-Hydro} for additional details). As a consequence the Lorentz force density on charge flow, $\boldsymbol{l} \equiv \rho \mathbf{E}  + \mathbf{J} \times \mathbf{B}$, decomposes into 
$\boldsymbol{l} = \rho \boldsymbol{L}/e + \mathbf{j} \times \mathbf{B}$, and $\boldsymbol{L} = e \left( \mathbf{E}  + \mathbf{u} \times \mathbf{B} \right)$, where $\mathbf{E}$ is an in-plane electric field and $\mathbf{B} = B \hat{\mathbf{z}}$ is a transverse magnetic field. The carrier imbalance flux $\mathcal{I}$ captures the electron-hole generation/recombination processes due to higher-order Coulomb collisions [\onlinecite{Foster2009,Naro2015,Lucas2018,Alekseev2018}] and electron-optical phonon scatterings [\onlinecite{Xie2016}]. The dissipation power density $\varpi$ and friction force density $\boldsymbol{f}$ describe the energy and momentum relaxations induced by phonon and impurity scatterings. We assume that phonons serve as an infinitely large thermal reservoir and define the global equilibrium while the carrier temperature fluctuations are allowed due to finite cooling.


\begin{figure}
\includegraphics[width=0.235\textwidth]{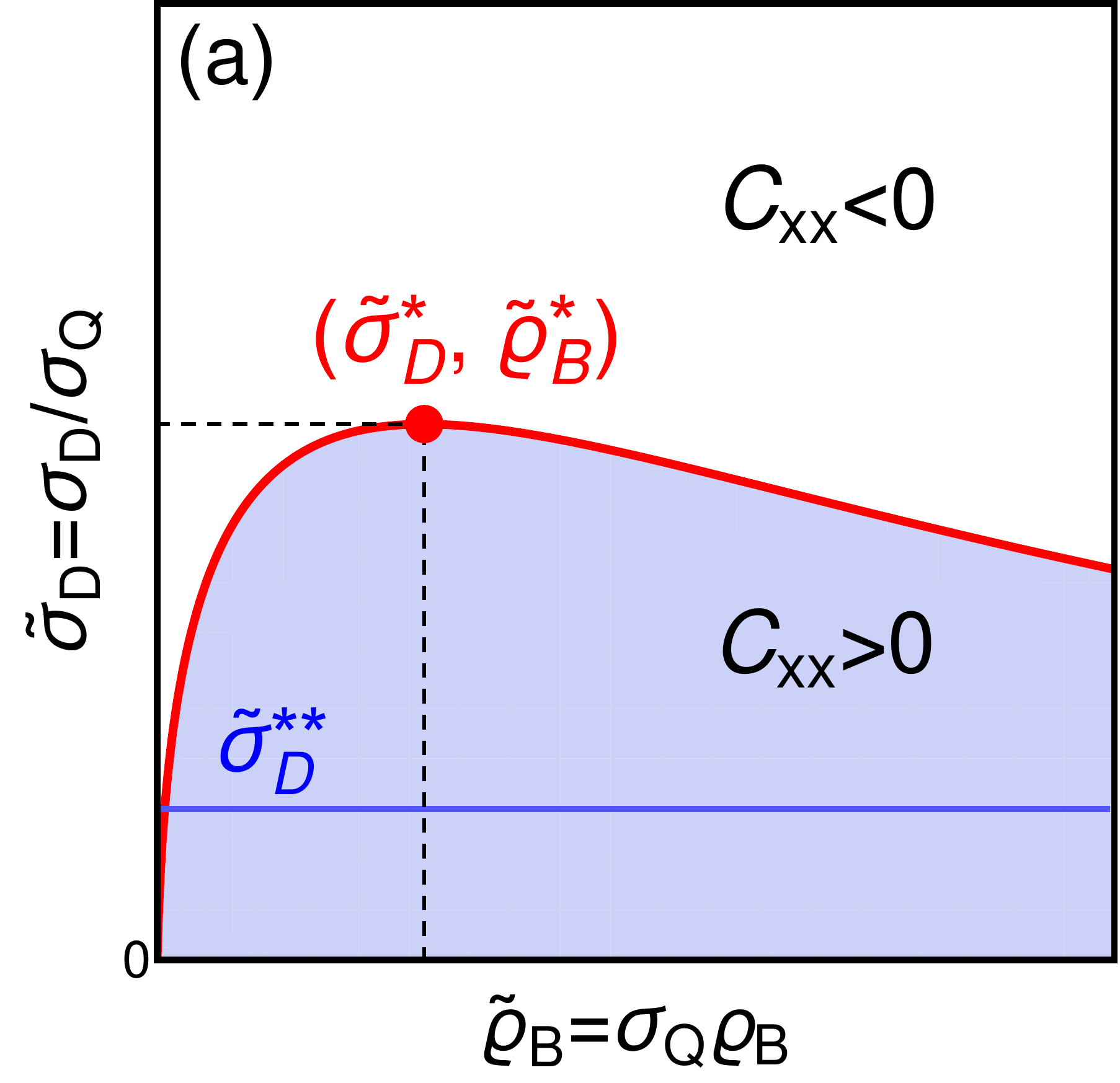}
\includegraphics[width=0.235\textwidth]{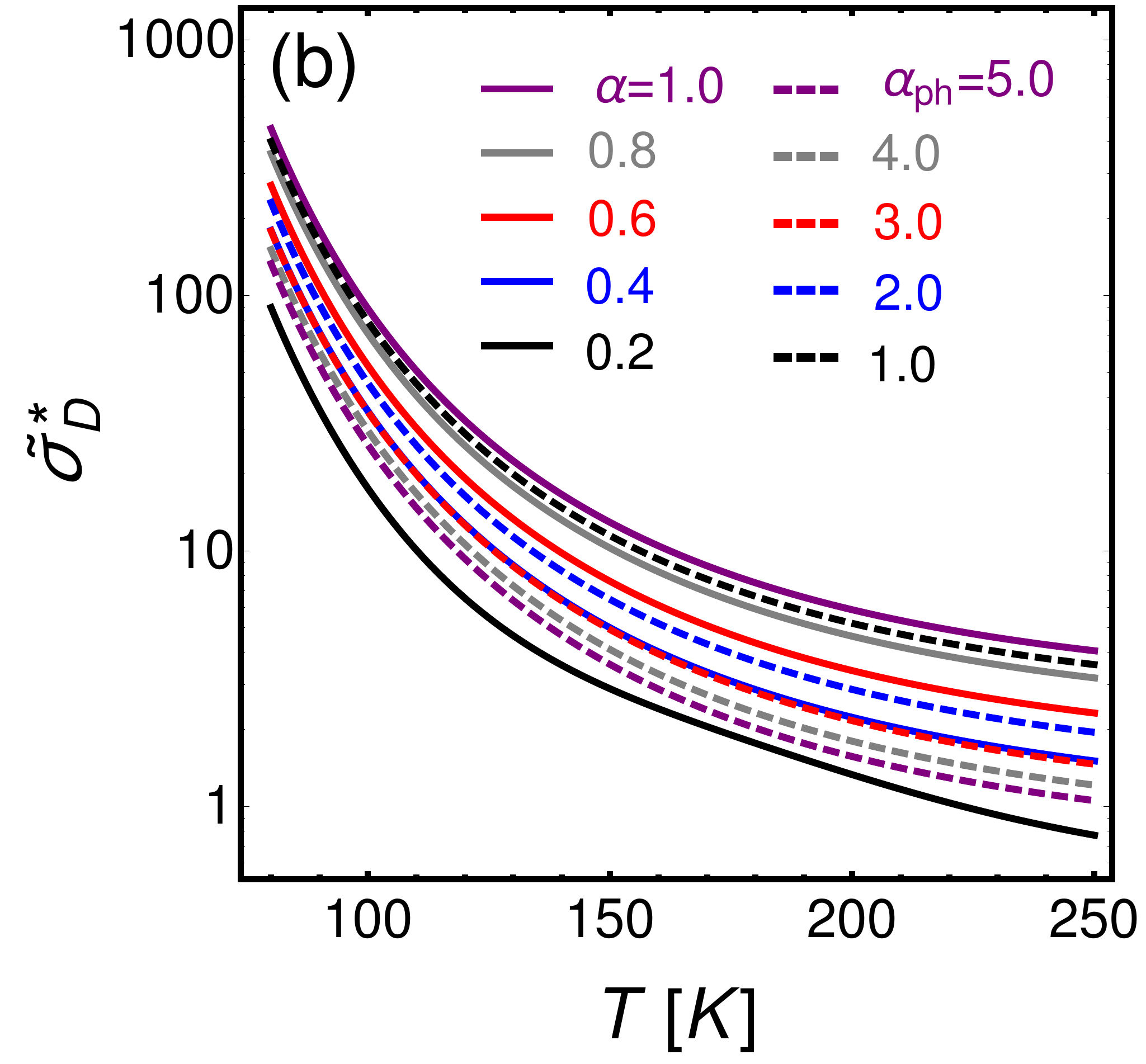} \\
\includegraphics[width=0.235\textwidth]{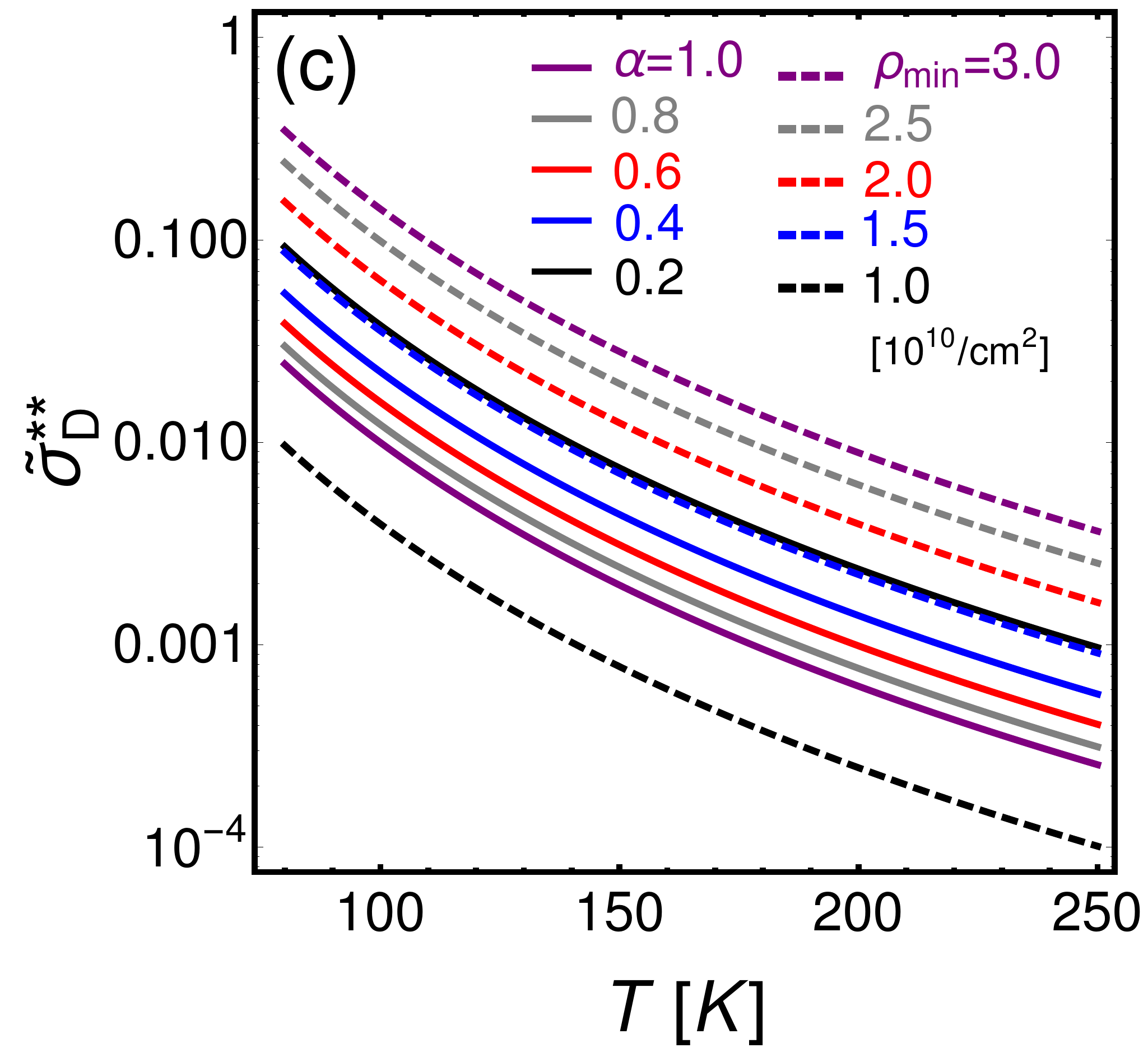}
\includegraphics[width=0.235\textwidth]{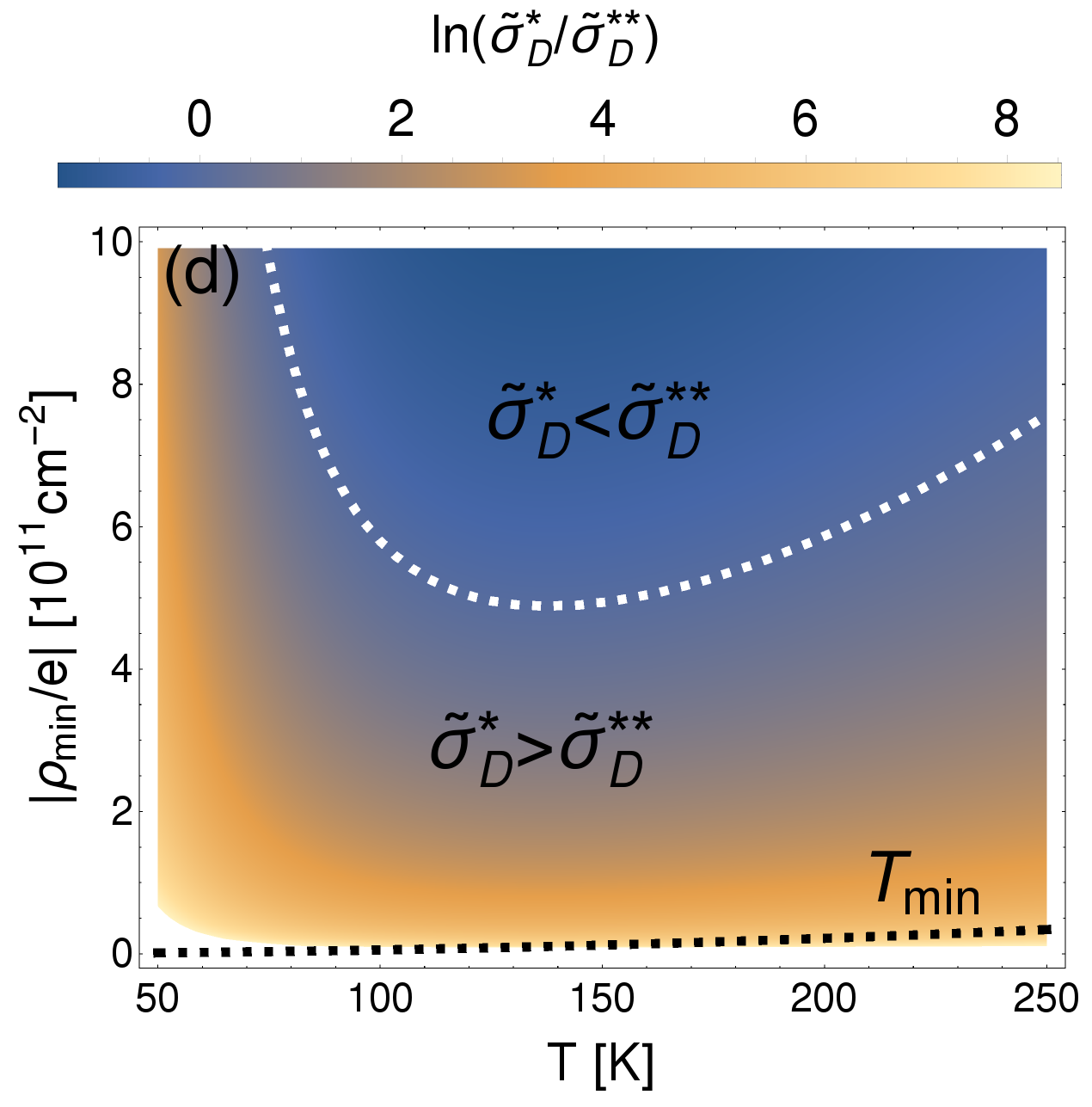} \\
\includegraphics[width=0.235\textwidth]{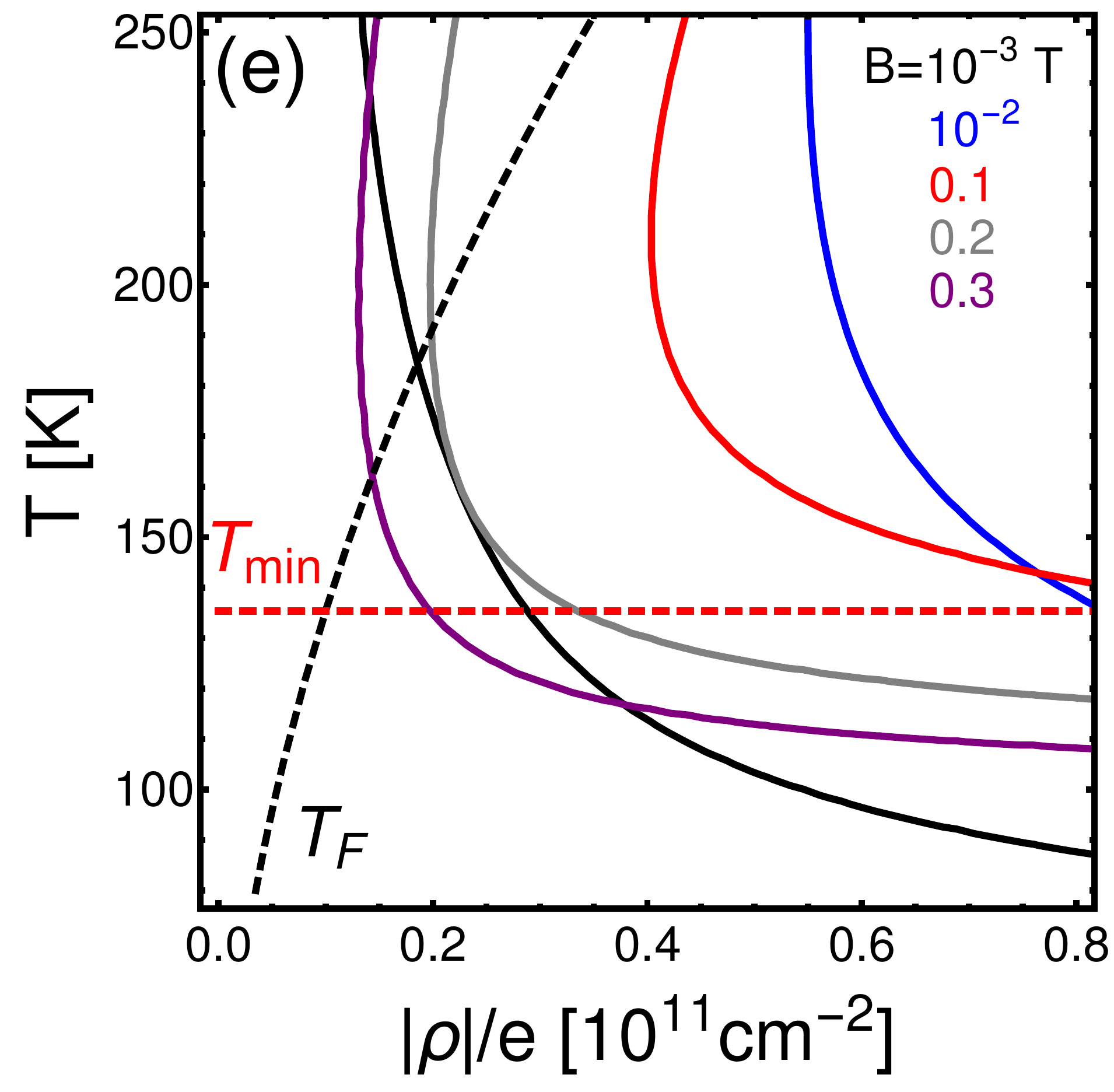}
\includegraphics[width=0.235\textwidth]{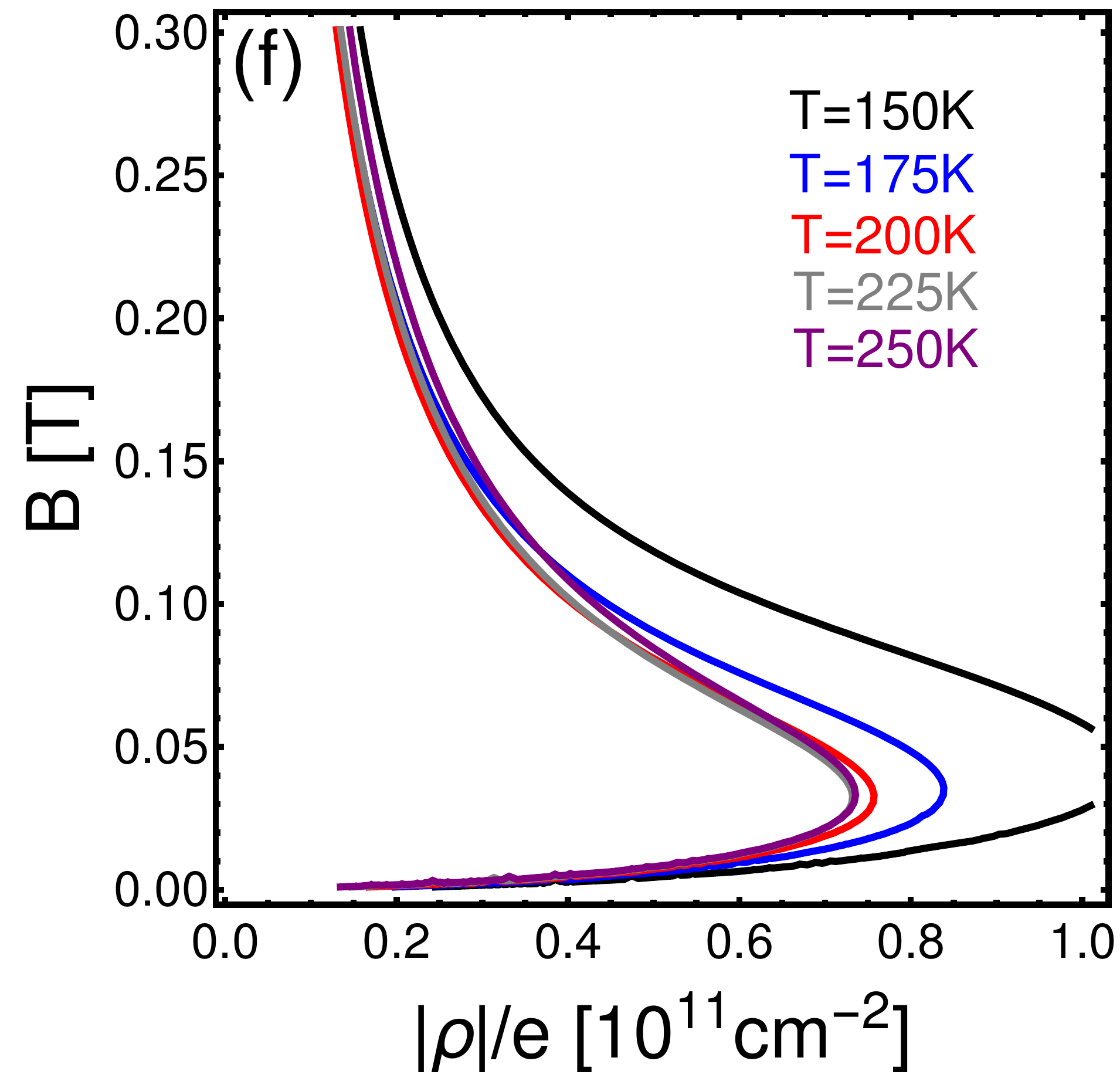}
\caption{Phase diagram of the effective viscosity. (a) The sign of $C_{xx}$ as a function of the effective Drude resistivity $\td{\s}_{D}$, magnetoresistivity $\td{\varrho}_B$, and imbalance-viscosity ratio $r$. The red curve $\td{\s}_{D}(\td{\varrho}_B;r)$ indicates $C_{xx}=0$, with the maximum $\td{\s}_{Q}^{\ast}$ at $\td{\varrho}_B^{\ast}$. The blue line indicates the lower bound of the Drude conductivity $\td{\s}_D^{\ast\ast}$ estimated by the inelastic scattering time $\tau_{\mathrm{ee}}$ and minimum charge density $\rho_\mathrm{min}$. (b)--(d) $\td{\s}_D^{\ast}$ and $\td{\s}_D^{\ast\ast}$ as functions of temperature $T$ and residual charge density $\rho_\mathrm{min}$. For comparison, we took several different values for the fine-structure constant $\a$ and the effective electron-phonon coupling $\a_\mathrm{ph}$, whose range is shown in the legends of (b) and (c). To obtain the phase diagram in (d) we used $\a = 0.6$ and $\a_\mathrm{ph} = 2.2$. The black dashed line indicates the puddle temperature $T_\mathrm{min}$. Panels (e) and (f)  show the sign of $C_{xx}$ as a function $\rho$, $T$, and $B$. The curves indicate $C_{xx}=0$ and the low-density region $C_{xx} >0$. We take the residual charge density $|\rho_\mathrm{min}| = 5 \times 10^{9} \, e/\mathrm{cm}^{2}$ ($T_\mathrm{min} = 135\,\mathrm{K}$). (e) The sign of $C_{xx}$ as a function $\rho$, $T$ for various $B$. The red and black dashed lines indicate the puddle temperature $T_\mathrm{min}$ and the Fermi temperature $T_F$, respectively. (f) The sign of $C_{xx}$ as a function $\rho$, $B$ for various $T$.} 
\label{fig1}
\end{figure}


\section{Linear transport theory} \label{Sec-Linear-Hydro} 

Within the linear response, the entropy production equation \eqref{en-prod} implies that the thermodynamic forces $\{\mathbf{L} + T \nabla(\mu/T), T \nabla(\nu/T), \nu, \del{T}, \mathbf{u}, \nabla \mathbf{u}\}$ determine the conjugate dissipative fluxes $\{\mathbf{j}, \mathbf{p}, \mathcal{I}, \varpi, \boldsymbol{f}, \hat{\theta} \}$ via the linear matrix relations: 
\begin{subequations} \label{linear-resp}
\begin{align}
& \begin{pmatrix}
\mathbf{j} \\ e \mathbf{p}
\end{pmatrix} = 
\begin{pmatrix}
\s_{00} & \s_{01} \\
\s_{10} & \s_{11} 
\end{pmatrix} \begin{pmatrix}
\mathbf{L}/e + T \nabla( \mu/e T ) \\ -T \nabla(\nu/e T )
\end{pmatrix}, \label{sigmas} \\
& \begin{pmatrix} e^2\mathcal{I} \\ \frac{e^2 \varpi}{T} \end{pmatrix} = -\begin{pmatrix}
\l_{11}   &   \l_{12} \\
\l_{21}   &   \l_{22}  
\end{pmatrix} \begin{pmatrix} \nu \\ \del T \end{pmatrix}, \label{lambdas} \\
& \boldsymbol{f} = - h \mathbf{u}/(v_F^{2}\tau_\mathrm{el}), \label{taus} \\
& \hat{\theta} = -\eta \left( \nabla \mathbf{u} + \nabla \mathbf{u}^\trasp \right) - \mathbb{I}  (\zeta - \eta)  \mathrm{div}\,\mathbf{u}. \label{viscs}
\end{align}
\end{subequations} 
Onsager's reciprocity enforces the symmetry of the kinetic coefficients: $\s_{\a\b} = \s_{\b\a}$, $\a,\b \in \{0,1\}$, and $\l_{\a\b}=\l_{\b\a}$, $\a, \b \in \{1,2\}$. In Eq.~\eqref{sigmas} the electric conductivities $\big\{\s_{\a\b}\big\}$ arise solely due to Coulomb collisions, which are functions of the dimensionless variables $\big\{\frac{\mu}{T}, \frac{\nu}{T} \big\}$. Particle-hole symmetry requires that the diagonal and off-diagonal elements are even and odd functions of the relative chemical potentials, respectively, $\s_{\a\b}(-\mu)= (-1)^{\a+\b}\s_{\a\b}(\mu)$. This implies that at local charge neutrality $\s_{01}(0) = 0$. In Eq.~\eqref{lambdas} the parameters $\{\l_{\a\b}\}$ characterize the efficiency of electron-hole generation/recombination and energy relaxation processes. In Eq.~\eqref{taus} the friction force density is determined by the momentum relaxation time $\tau_\mathrm{el}$ caused by impurities and phonons. In Eq.~\eqref{viscs} $\eta$ and $\zeta$ are, respectively, the shear and bulk viscosities. These kinetic coefficients can be computed via microscopic quantum kinetic equations [\onlinecite{Foster2009,Naro2015,Lucas2018,Xie2016,Alekseev2018,Ho2018}]. In particular, we compute matrix $\hat{\lambda}$ in Eq. \eqref{lambdas} along with impurity- and phonon-mediated relaxations in Appendix \ref{App-Relaxation}. We note that $\{ \mathcal{I}, \varpi, \boldsymbol{f} \}$ are present already at the level of the ideal hydrodynamics that omits the dissipative fluctuations. In contrast, the conductivities and viscosities $\{\s_{\a\b},\eta,\zeta\}$ require solving the kinetic equations in first order in $\tau_{\mathrm{ee,eh}}$. In what follows we assume that the response coefficients in Eq.~\eqref{linear-resp} are spatially uniform and magnetic field independent. These simplifications are justified in the linear response regime and for weak magnetic field $\w_B \tau_\mathrm{ee,eh} \ll 1$, where the cyclotron frequency of a quasiparticle is $\w_B = v_F^2 e n B/h$.
 
For discussing the general linear response we define the heat current density 
$ \mathbf{Q} \equiv h \mathbf{u} - \nu \mathbf{P} + \frac{\mu}{e} \mathbf{J}$, which substitutes for the fluid velocity $\mathbf{u}$ as an independent flow mode, and the electrochemical potential fluctuations $\del V$ through $-\nabla \del V = \boldsymbol{\mathcal{E}} \equiv \mathbf{E} + \nabla{\del \mu}/e$, where $\del{\mu}$ denotes the local chemical potential fluctuations. Combining Eqs.~\eqref{hyd-eqs} and \eqref{linear-resp}, we obtain the transport equations for steady flows in the form 
\begin{subequations} \label{trans-eq}
\begin{align}
& \nabla \cdot \boldsymbol{\Psi}_\a =  - \l_{\a \b} \Phi_\b, \label{cont-eq} \\
& \hat{\Pi}_{\a\b} \cdot  \boldsymbol{\Psi}_\b - H_{\a\b} \nabla^2 \boldsymbol{\Psi}_\b  = - Z_{\a\b} \nabla \Phi_\b. \label{b-l-resp}
\end{align}
\end{subequations} 
Here $\Phi_{\a}(\bfr) \in \big\{ \del{V}, \frac{1}{e} \nu, \frac{1}{e} \del{T} \big\}$ are the hydrodynamic potentials and $\boldsymbol{\Psi}_\a(\bfr) \in \big\{ \mathbf{J}, e \mathbf{P}, \frac{e}{T} \mathbf{Q} \big\}$ are the conjugate currents, with $\a \in \{0,1,2\}$ labeling the corresponding charge, carrier, and thermal modes, respectively. 
In Eq.~\eqref{cont-eq} the coefficients $\l_{\a\b}$ are the relaxation parameters in Eq.~\eqref{lambdas} complemented by $\l_{\a\b}=0$ for $\a$ or $\b=0$ (charge conservation). We have neglected the nonlinear Joule and viscosity-induced heating terms in the thermal continuity equation. In Eq.~\eqref{b-l-resp}, the bulk transport coefficients are given by
$\hat{\Pi}_{\a\b} = \left[ (h/v_F^{2}\tau_\mathrm{el}) \chi_\a \chi_\b + \varrho_{\delta \g} A_{\a \delta} A_{\b \g}\right] \hat{\mathbb{I}} + B_{\a\b} \hat{\mathbb{B}}$. Here $\chi_{\a} \in  \big\{ \frac{\mu}{eh}, \, 0 , \, -\frac{T}{eh} \big \}$, $\varrho_{\a \b} = [\hat{\s}^{-1}]_{\a\b}$ for $\a,\b \in \{0,1\}$, and $\varrho_{\a \b}=0$ for $\a$ or $\b=2$, encoding the electric conductivities in Eq.~\eqref{sigmas}; and $A_{\a \b} = \delta_{\a \b} + \chi_\a d_\b$, with $d_\b \in \big\{\rho, e n, 0 \big\}$. The magnetic field effects are described by the last term, where $\hat{\mathbb{B}}=B \hat{\ep}$, with $\hat{\ep}$ being the two-dimensional Levi-Civita symbol and $B_{\a\b} = B_{\b\a}$, with $B_{00} =  \chi_0 ( 2 +  \chi_0 \rho)$, $B_{02} =  \chi_2 ( 1 +  \chi_0 \rho)$, $B_{22} = (\chi_2 )^2 \rho$, and $B_{\a \b} =0$ for $\a=1$. The shear and bulk viscous effects are respectively described by $H_{\a\b} = \eta \chi_\a \chi_\b$, and  $Z_{\a\b} = \delta_{\a\b} +\zeta \chi_\a \chi_\g \l_{\g \b}$.
             

\begin{figure*}
\includegraphics[width=0.47\textwidth]{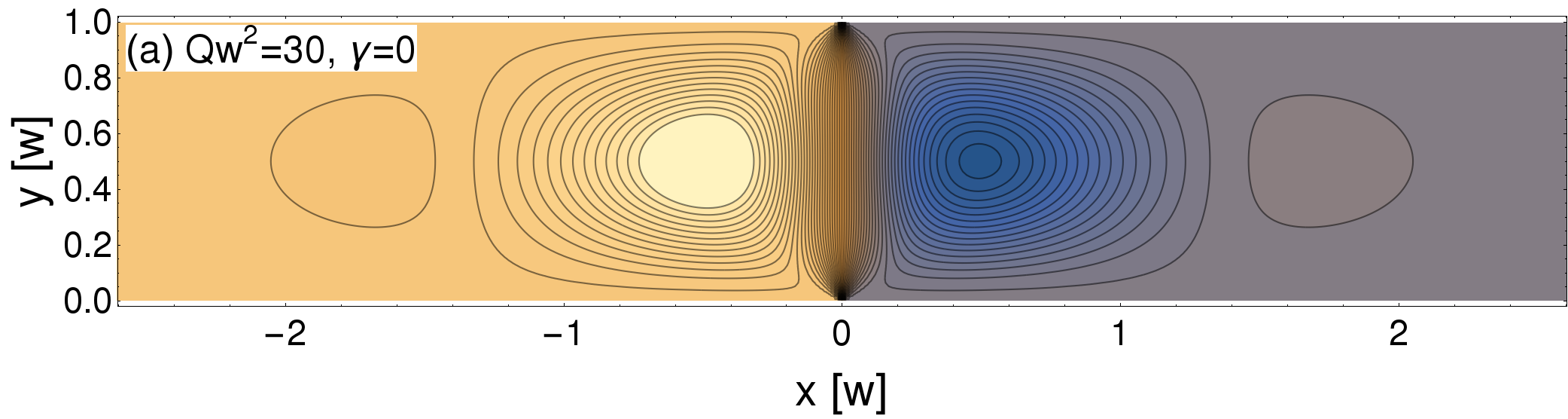} 
\includegraphics[width=0.47\textwidth]{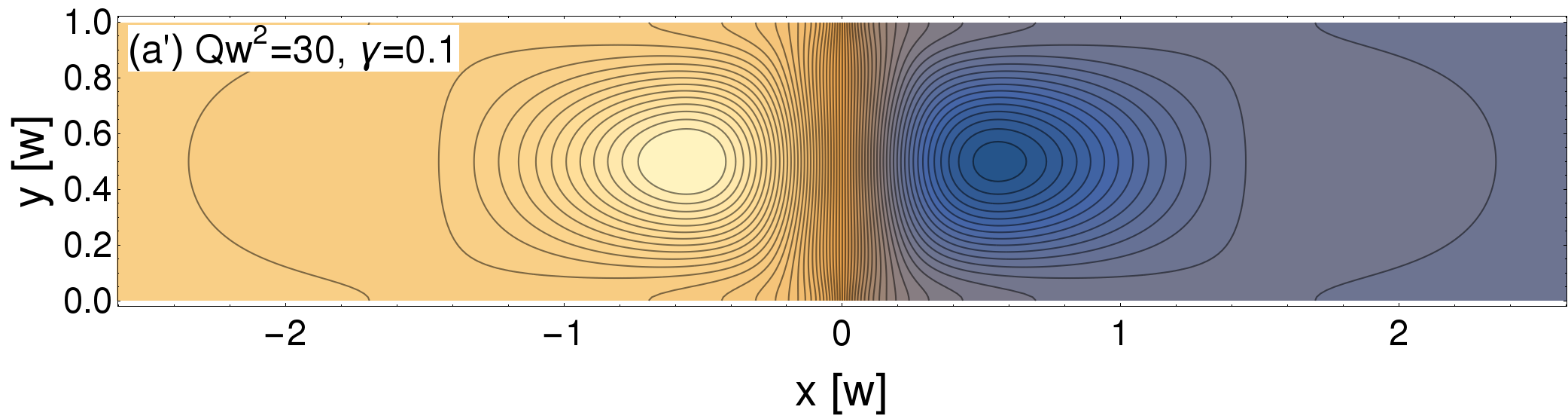} \\
\includegraphics[width=0.47\textwidth]{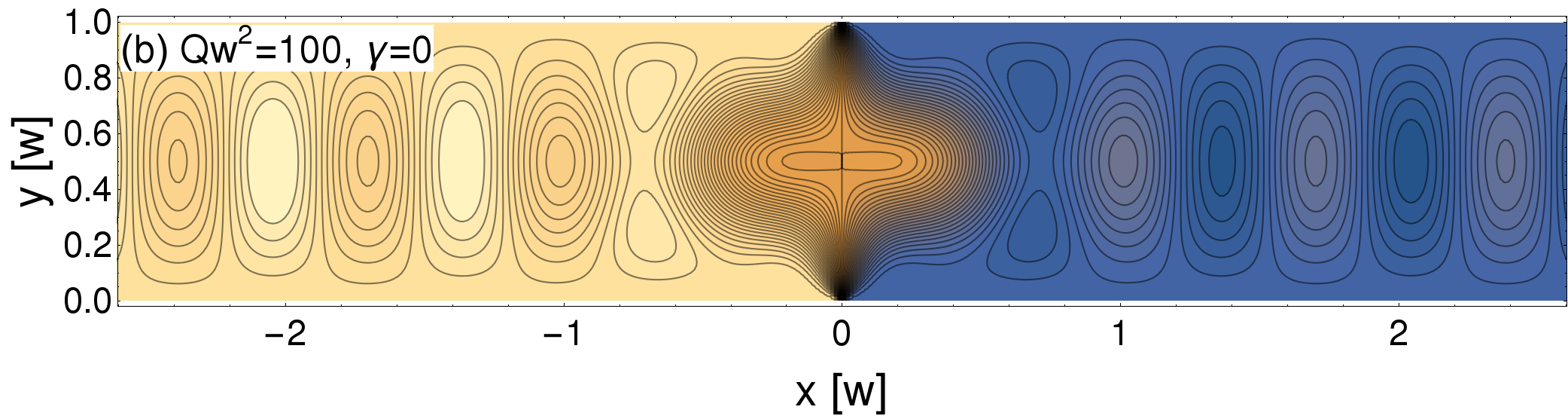} 
\includegraphics[width=0.47\textwidth]{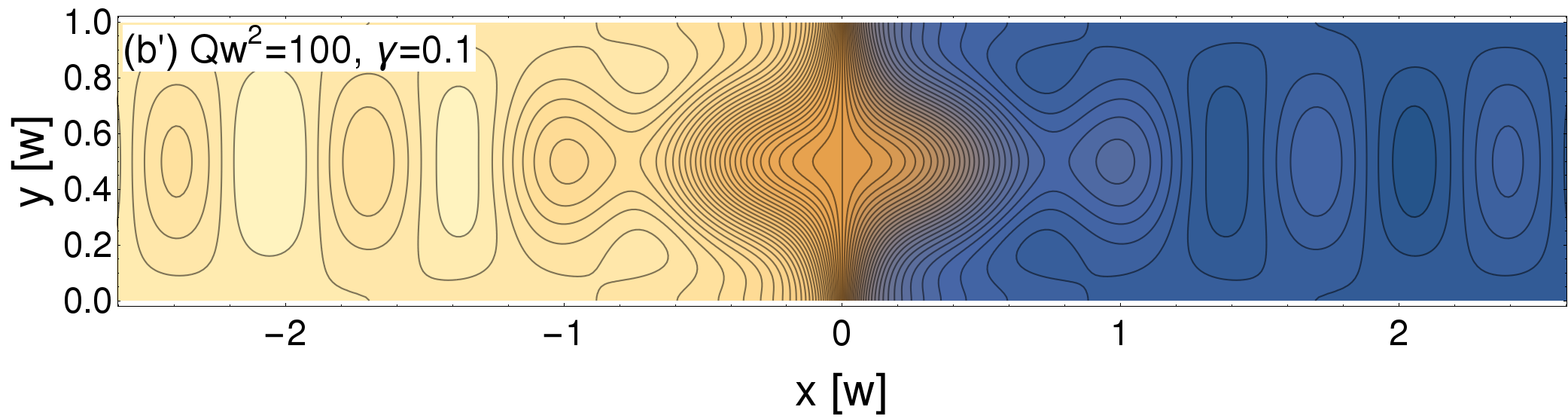} \\
\includegraphics[width=0.47\textwidth]{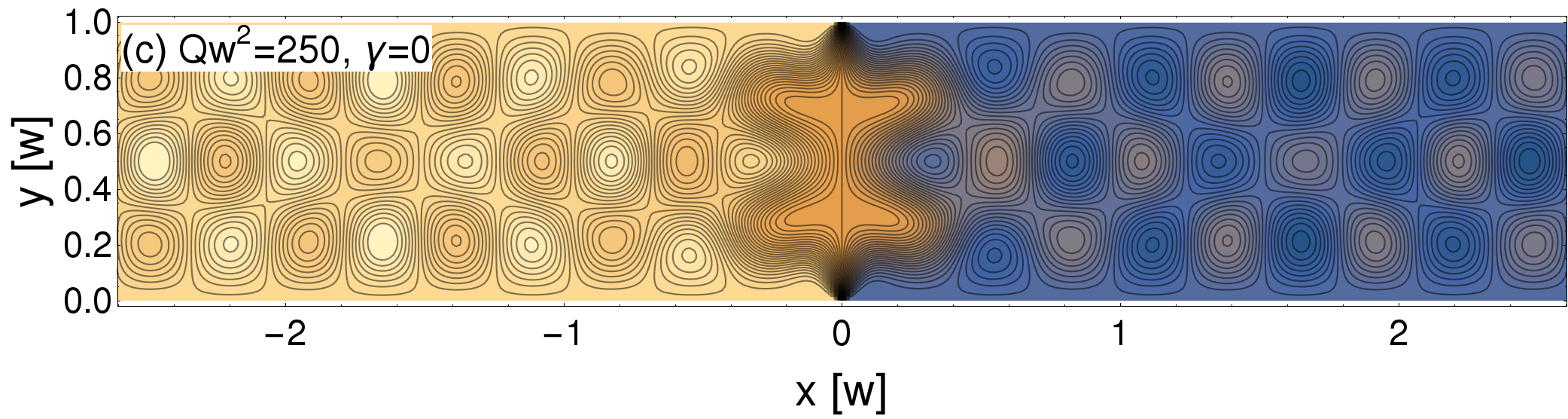} 
\includegraphics[width=0.47\textwidth]{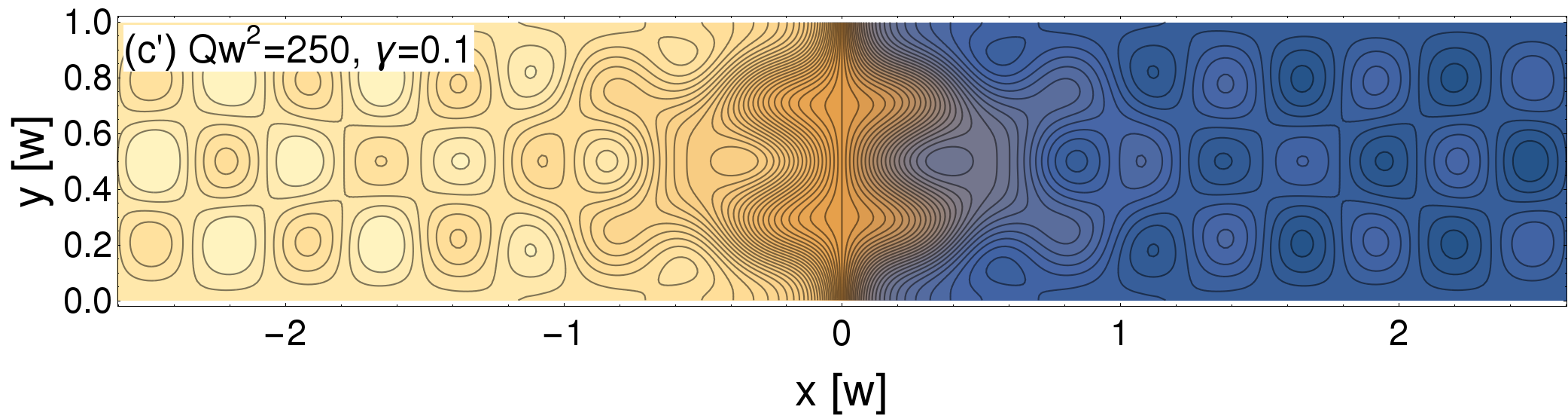}
\includegraphics[width=0.47\textwidth]{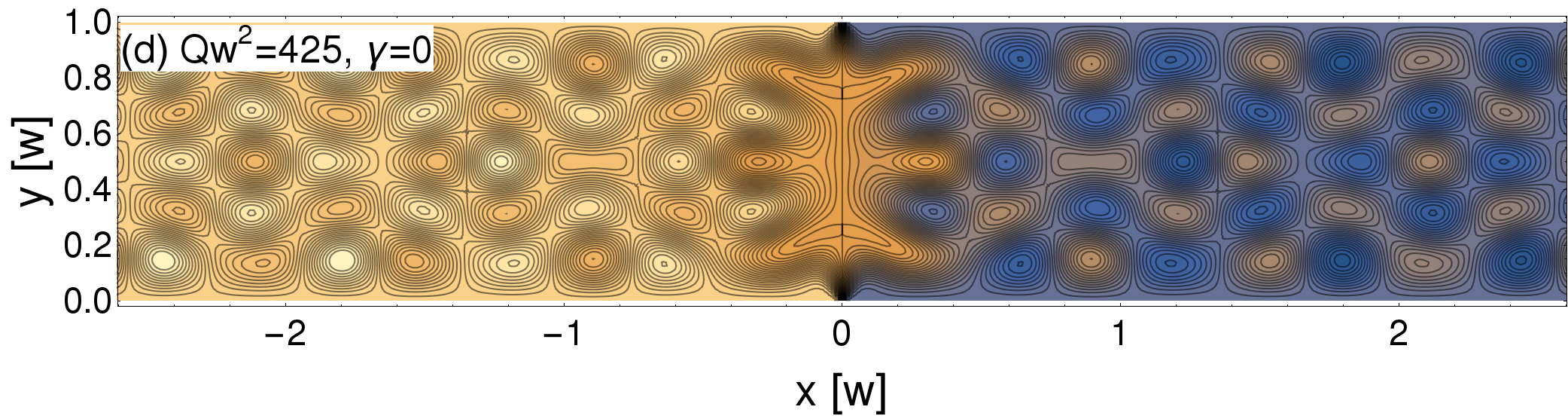} 
\includegraphics[width=0.47\textwidth]{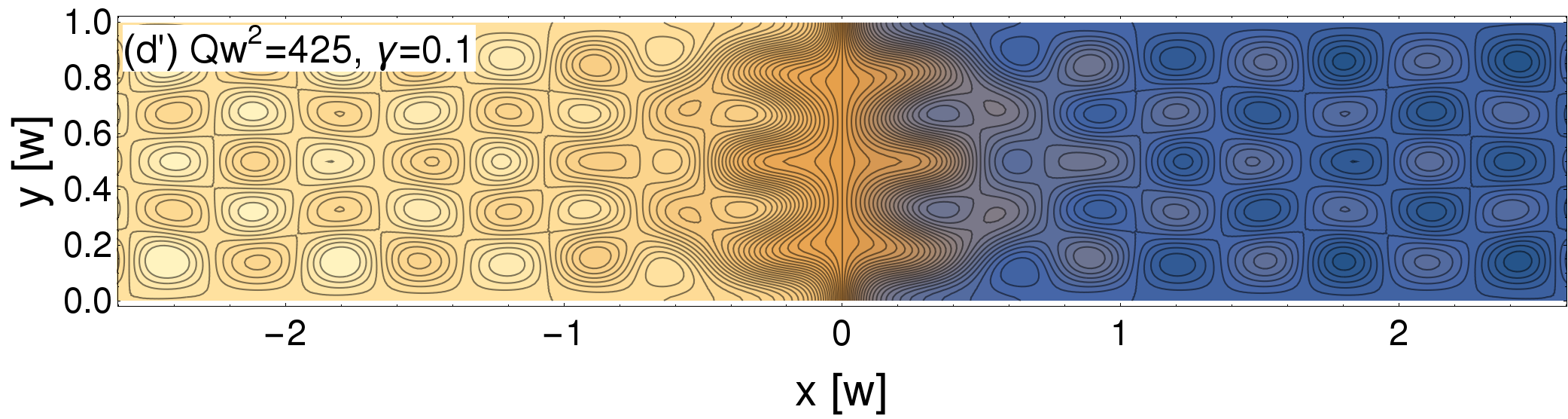} \\
\caption{Charge current distributions in a strip of monolayer graphene for various resistivity-to-viscosity-imbalance ratios $Q$ and lead sizes $\g$. We take the strip width $w=1$. (a)--(d) Point-like leads $\g=0$ and (a$^\prime$)-(d$^\prime$) finite-size leads $\g =0.1$. For $Q < Q^\ast$ the amplitudes of vortices decay rapidly with distance to leads. Moreover, only the first pair of vortices survives for finite lead width. For $Q > Q^\ast$ the vortices form eddy flows that are stable against the finite lead width.}    
\label{fig2}
\end{figure*} 


\subsection{Stream function of the charge flow} \label{Sec-Stream}

We focus on the electric response of the system. For that purpose we introduce the stream function for the charge flow $\varphi(\bfr)$ via $(J_{x}, J_{y}) \equiv (\d_y,-\d_x)\varphi$ to solve the charge conservation. Eliminating the carrier and thermal modes, we cast Eq.~\eqref{trans-eq} in the form of Ohm's law together with the stream function equation,
\be \label{E-response}
\boldsymbol{\mathcal{E}} = \hat{\mathcal{R}} \cdot \mathbf{J}, \quad \hat{\mathcal{R}} = \hat{\mathcal{L}}_{00} - \hat{\mathcal{G}}_{0 \mathrm{a}} \hat{\mathcal{K}}_{\mathrm{a} \mathrm{b}} \hat{\mathcal{L}}_{\mathrm{b} 0}, \quad \mathcal{R}_{xx} \nabla^2 \varphi=0,  
\ee
where the summation of mode indices spans only $\{\mathrm{a},\mathrm{b}\} \in \{1,2\}$; $\hat{\mathcal{L}}_{\a \b} = \hat{\Pi}_{\a \b} - H_{\a \b} \hat{\mathbb{I}} \nabla^2$; $\hat{\mathcal{G}}_{\a \mathrm{b}} = \hat{\mathcal{L}}_{\a \mathrm{b}} - \mathsf{Z}_{\a \mathrm{b}} \hat{D}$, where $\hat{D} \equiv \nabla \nabla$ is the gradient-gradient tensor
and $\mathsf{Z}_{\a \mathrm{b}} = \del_{\a \mathrm{c}} [\hat{\l}^{-1}]_{\mathrm{c} \mathrm{b}}+ \zeta \chi_\a \chi_\mathrm{b}$; and $[\hat{\mathcal{K}}^{-1}]_{\mathrm{a} \mathrm{b}} = \hat{\mathcal{G}}_{\mathrm{a} \mathrm{b}}$ is an analog of the propagator in carrier-thermal-mode space. We note that the second term in the resistivity operator $\hat{\mathcal{R}}$ encodes the thermoelectric and imbalance-electric effects, through which the imbalance relaxation and bulk viscous processes proliferate. In practice one first solves the boundary value problem for the stream function $\varphi(\bfr)$ to obtain the charge flow $\mathbf{J}(\bfr)$. The relative voltage between two space points $\bfr_1$ and $\bfr_2$ takes the integral form $\del{V}(\bfr_1) - \del{V}(\bfr_2) = \int_{c} d \mathbf{l} \cdot \hat{\mathcal{R}} \cdot \mathbf{J}$, where ``$c$'' denotes a path from $\bfr_1$ to $\bfr_2$ and $d \mathbf{l}$ is an infinitesimal vector element of the path. It is obvious that the viscosities and imbalance relaxation processes contribute to the electrochemical field only if the charge flow is inhomogeneous.  

The charge conservation imposes important constraints on electric responses. (i) Finite imbalance relaxation and bulk viscosity encoded in $\mathsf{Z}_{\a \mathrm{b}}$ contribute only to the magnetoresponses, unlike the shear viscosity that contributes at $B=0$.  (ii) The resistivity operator $\hat{\mathcal{R}}$ is effectively a function of the Laplacian $\nabla^2$ even though $\hat{D}$ involves more types of derivative operations. (iii) $\hat{\mathcal{R}}_{xx}=\hat{\mathcal{R}}_{yy}$ and $\hat{\mathcal{R}}_{xy}=-\hat{\mathcal{R}}_{yx}$ are even and odd functions of the magnetic field $B$, respectively. These properties can be readily proved by formally expanding $\hat{\mathcal{R}}$ in a Dyson series of $\hat{D}$ and applying the constraints $\hat{D} \cdot \mathbf{J} =0$ and $\hat{D}\hat{\ep}\cdot \mathbf{J} = \nabla^2\hat{\ep}\cdot\mathbf{J}$ (see Appendix \ref{App-R} for further details).

We benchmark our theory in two limits. The first is Ohmic flow. For inviscid and post balanced fluids, $\eta,\zeta,\hat{\l}^{-1}=0$, the resistivity operator in Eq. \eqref{E-response} reduces to the bulk resistances, $\hat{\mathcal{R}} \to \hat{R}$; the stream function satisfies the harmonic equation $\nabla^2 \varphi =0$, and no vertex is allowed. The second is the Stokes flow considered in Refs. [\onlinecite{Levitov}] and [\onlinecite{Polini}]. At zero field $B=0$ and in the shear-viscosity-dominant regime, the resistivity operator reduces to Laplacian $\hat{\mathcal{R}} \sim \nabla^2$, and the stream function satisfies the biharmonic equations $\nabla^4 \varphi =0$.

For weak inhomogeneity we expand the resistivity operator up to first order in $\nabla^2$, $\hat{\mathcal{R}} \simeq \hat{R} + \del{\hat{R}} + \mathcal{O}(\nabla^4)$ (see Appendix \ref{App-Expansion} for the intermediate steps of calculation), where the imbalance-viscosity corrections read $\del \hat{R} = ( C_{xx} \hat{\mathbb{I}} + C_{xy} \hat{\ep} ) \nabla^2$. The stream function equation reduces to
\be \label{stream-eq}
R_{xx} \nabla^2 \varphi + C_{xx}\nabla^4 \varphi=0,
\ee
where $R_{xx}(B)= \s^{-1}[1+c_0(B)]$ and $\rho^2 C_{xx}(B) = -c_1(B) \eta + c_2(B) \left( \eta+\zeta+ \varsigma\right)$:
$\s = \s_Q + \s_D$ is the hydrodynamic conductivity, with $\s_Q=\s_{00}$ and $\s_D = v_F^2 \tau_\mathrm{el} \rho^2/h$ being the minimal and Drude conductivities, respectively, and $\varsigma = \kappa_\mathrm{a} \kappa_\mathrm{b} \hat{\l}_{\mathrm{a}\mathrm{b}}^{-1}$, with $\kappa_\mathrm{a} \in \big\{ e n - \rho \frac{\s_{01}}{\s_{00}}, \frac{eh}{T} \big\}$, representing the effective viscosity induced by imbalance relaxation processes. The dimensionless functions $c_0(B) = \s_Q \varrho_B/\Xi$, $c_1(B) = (R_{xx} \s_D)^2$, and $c_{2}(B) = \s_D \varrho_B/\Xi^2$, where $\Xi = (1+\s_D/\s_Q)^2 + \s_D \varrho_B$ and $\varrho_B = v_F^2 \tau_\mathrm{el} B^2 /h$ is the magnetoresistivity at neutrality. At zero field $c_0(0)=c_2(0)=0$ and $c_{1}(0)=\s_D/\s$.

Equation~\eqref{stream-eq} is characterized by the effective resistivity-to-viscosity ratio $Q \equiv R_{xx}/C_{xx}$. We note that the sign of $Q$ is not fixed by any fundamental reason. In particular, at neutrality $\rho=0$, $C_{xx}= \left( \eta+\zeta+ \varsigma\right) (v_F^2 \tau_\mathrm{el}B/h)^2>0$. In contrast, for $\s_D/\s_Q \gg1$, $C_{xx} = -\eta/\rho^2 < 0$. $C_{xx}$ remains positive at low charge density for a moderate strength of the momentum relaxation scattering as captured by $\tau_\mathrm{el}$. As shown in Fig.~\ref{fig1}(a), the critical line $C_{xx}=0$ gives the equation for three effective quantities $\td{\s}_{D}(1+\td{\s}_{D}+\td{\varrho}_B)^2=(1+r)\td{\varrho}_B$, where $\td{\s}_{D} \equiv \s_{D}/\s_Q$, $\td{\varrho}_B \equiv \s_Q \varrho_B$, and $r \equiv (\zeta+\varsigma)/\eta$ are the dimensionless Drude resistivity, magnetoresistivity, and imbalance-viscosity ratio, respectively. The $C_{xx}>0$ regime is accessible as long as $\td{\s}_{D}^{\ast\ast} < \td{\s}_{D} < \td{\s}_{D}^{\ast}$, where the upper bound $\td{\s}_{D}^{\ast} = (\sqrt{r+2}-1)/2$ at $\td{\varrho}_B^{\ast} = (\sqrt{r+2}+1)/2$, determined by the imbalance-viscosity ratio $r$, and the lower bound $\td{\s}_D^{\ast\ast} \sim v_F^2 \rho_\mathrm{min}^2 \tau_{ee}/h$, estimated by the inelastic scattering time $\tau_\mathrm{ee(eh)}$ and the residual charge density $\rho_\mathrm{min}$. 

For high-quality hBN graphene close to neutrality, we numerically calculate the imbalance relaxation coefficient $\varsigma$ and the scattering time $\tau_\mathrm{el}$ using the kinetic theory in Appendix \ref{App-Relaxation} and estimate $\eta \approx 0.45 T^2/v_F^2 \a^2 $ [\onlinecite{Mueller2009,Principi2016,Sherafati2016}], $\tau_\mathrm{ee}^{-1} \approx \a^2T$ [\onlinecite{Mueller2008,Lucas2018}], $\s_Q \approx (0.79 + 9.13 \a)/\a^2$ [\onlinecite{Xie2016}], and $\s_{01},\zeta \approx 0$. We take the fine-structure constant $\a = 0.6$ and the effective electron-phonon coupling $\a_\mathrm{ph} = 2.2$ [\onlinecite{Xie2016}]. In Fig.~\ref{fig1} we show $\td{\s}_D^{\ast}$ and $\td{\s}_D^{\ast\ast}$ as functions of temperature $T$ and residual charge density $\rho_\mathrm{min}$ and observe that $\td{\s}_D^{\ast} \sim 10^4 \td{\s}_D^{\ast\ast}$ for $T>T_\mathrm{min}$, where the puddle temperature $T_\mathrm{min}\approx T_F(\rho_\mathrm{min})$ with $T_F(\rho)= v_F \sqrt{\pi |\rho| /e}$ being the Fermi temperature. We further show the sign of $C_{xx}$ in $\rho$, $T$, and $B$ space. We take $|\rho_\mathrm{min}| = 5.0 \times 10^{9} \, e/\mathrm{cm}^{2}$ [\onlinecite{Crossno}] so that $T_\mathrm{min} = 135\,\mathrm{K}$ and observe that at about $B \sim 0.1 \, \mathrm{T}$ and $T>T_\mathrm{min}$, we access the $C_{xx} >0$ region for $|\rho_\mathrm{min}| \lesssim |\rho| \lesssim 10^{11} \, e/\mathrm{cm}^2$.     

\begin{figure*}
\includegraphics[width=0.235\textwidth]{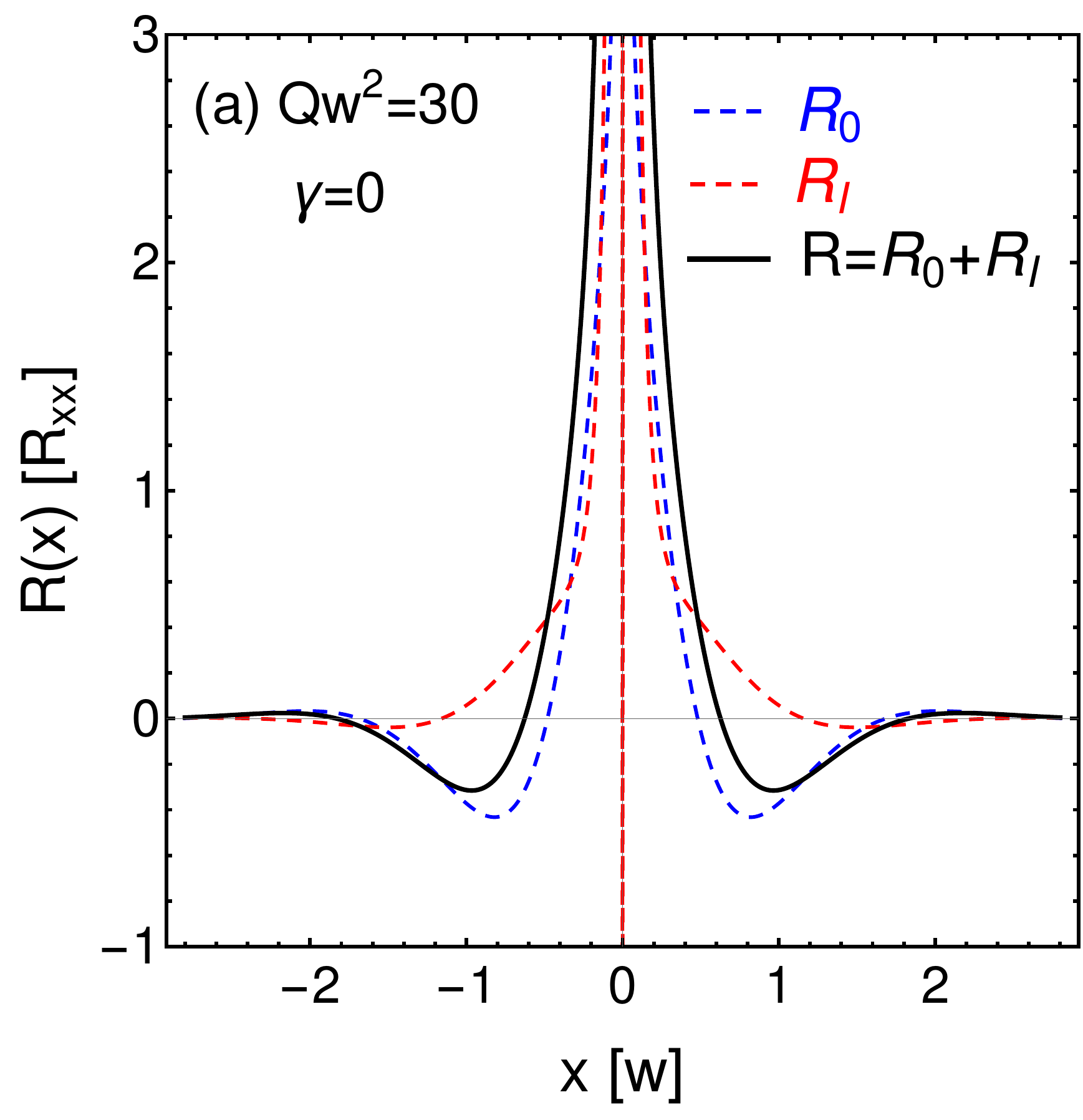} 
\includegraphics[width=0.235\textwidth]{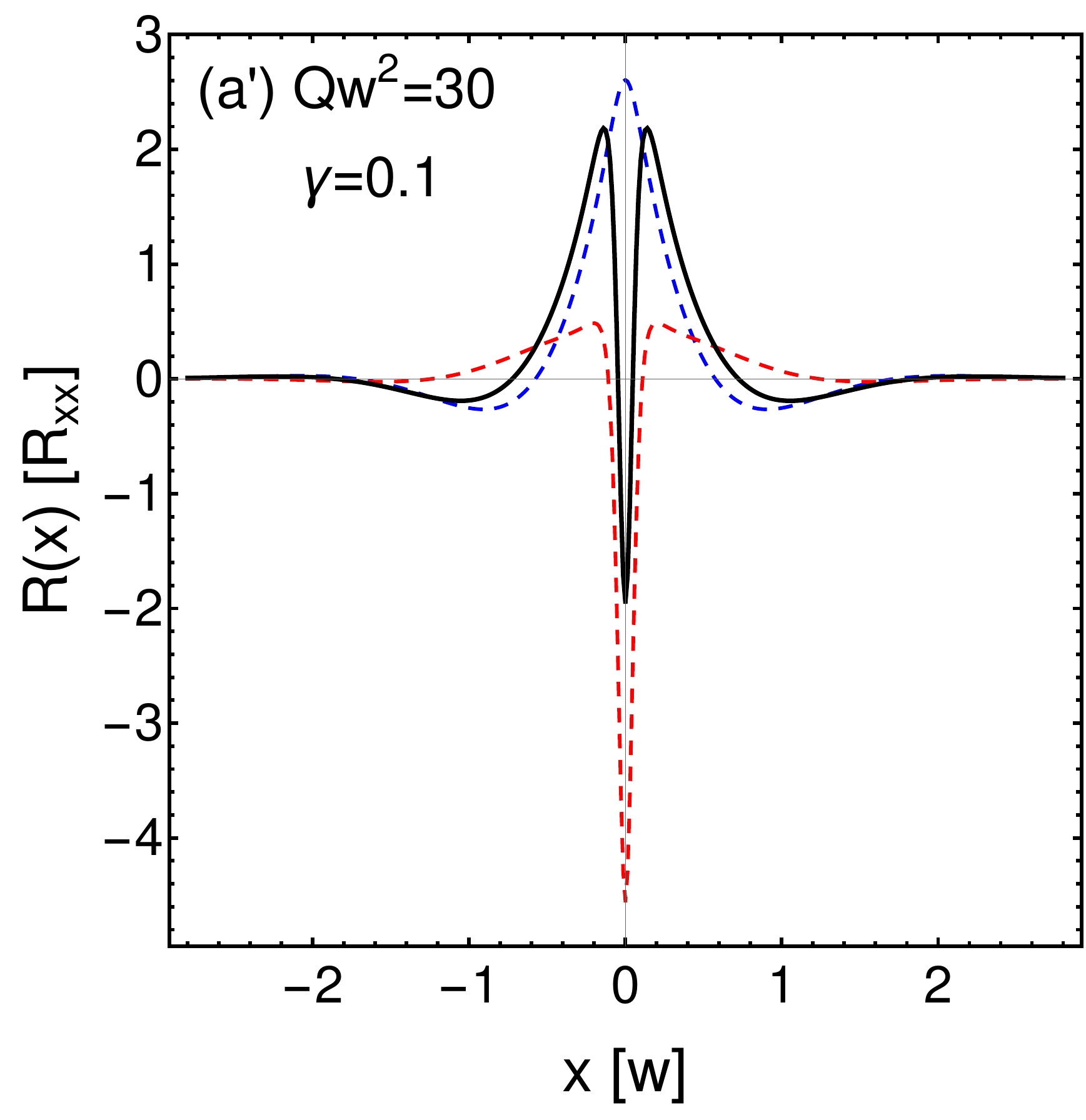}
\includegraphics[width=0.235\textwidth]{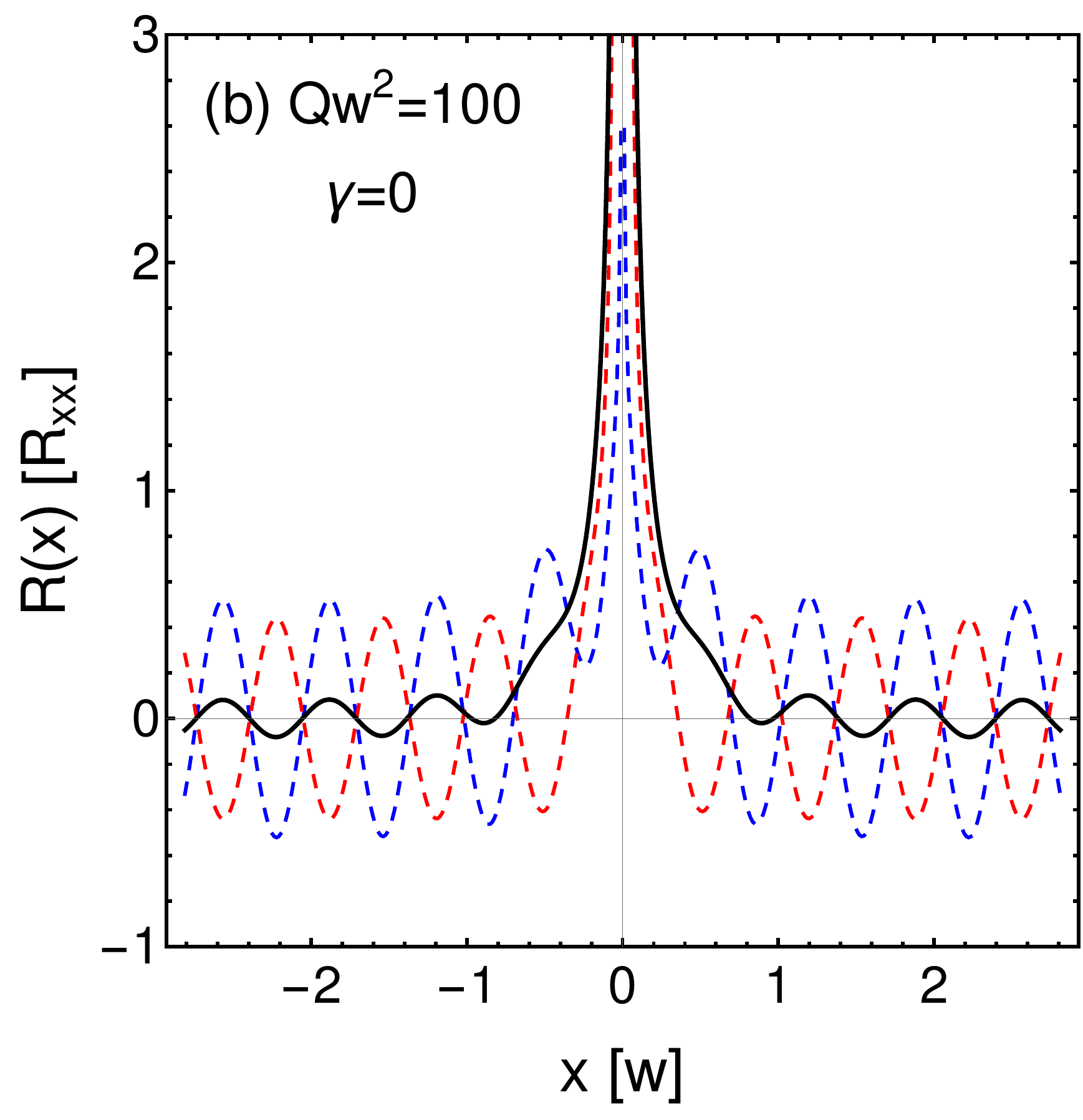} 
\includegraphics[width=0.235\textwidth]{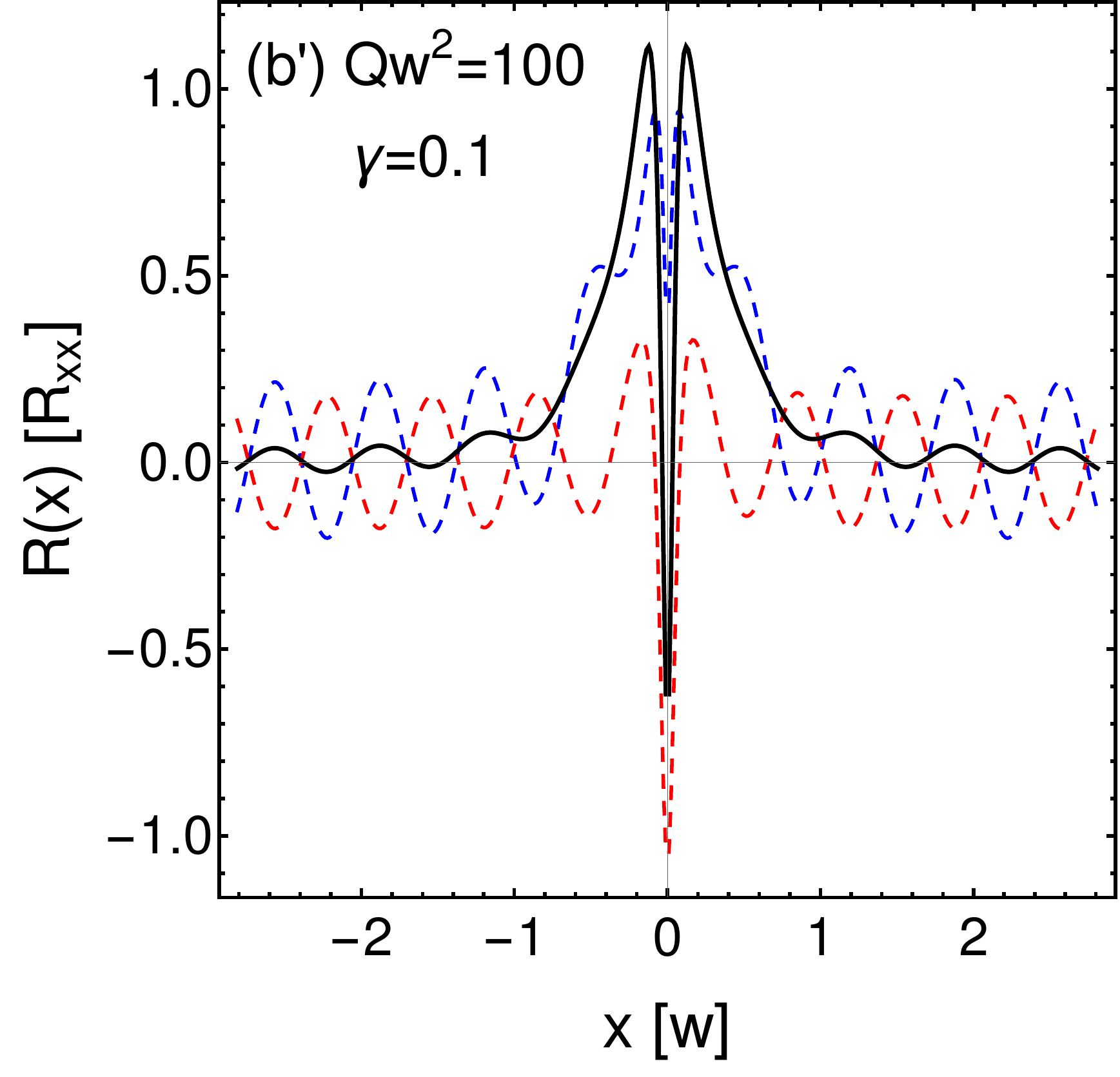} \\
\includegraphics[width=0.235\textwidth]{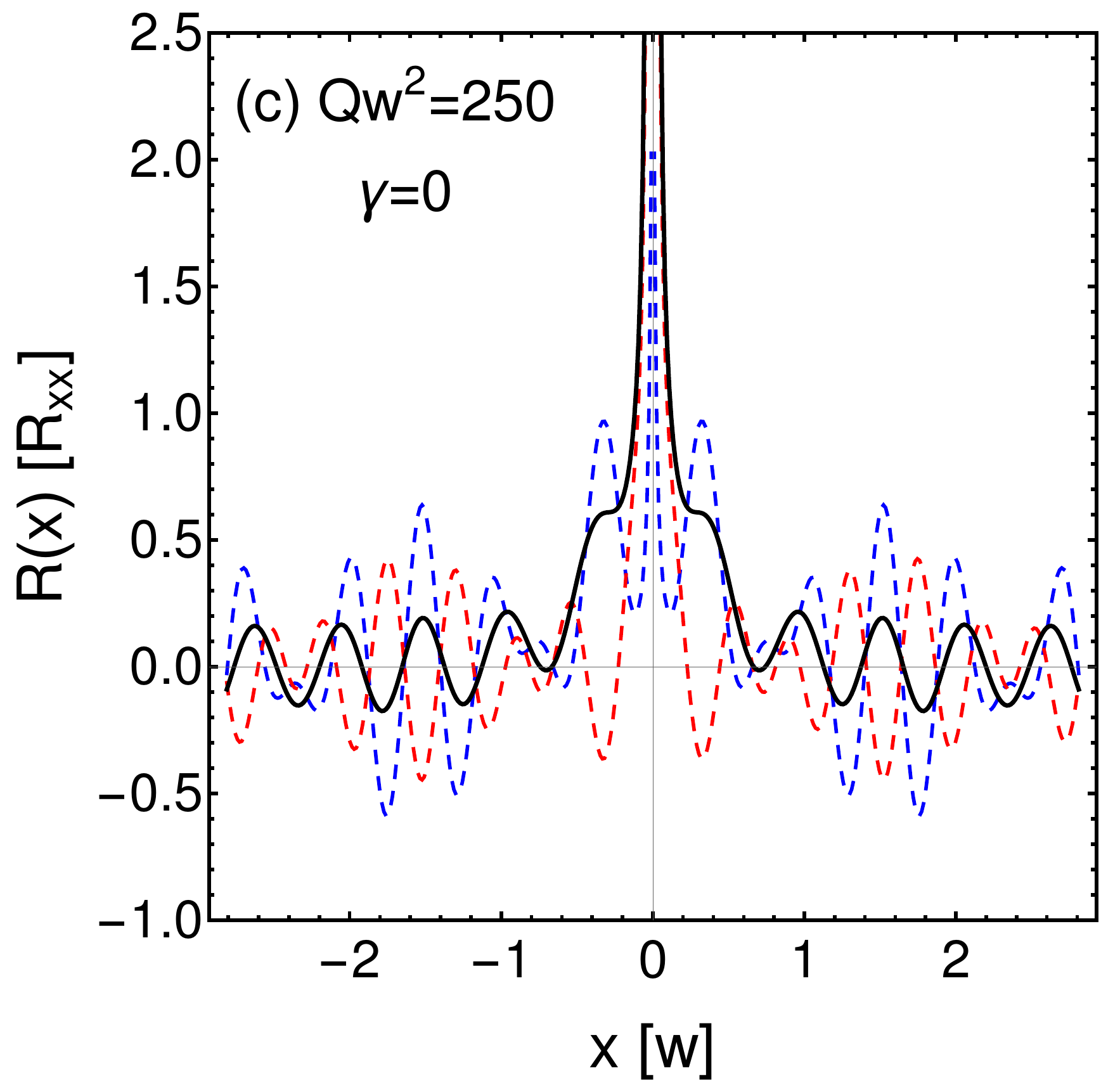} 
\includegraphics[width=0.235\textwidth]{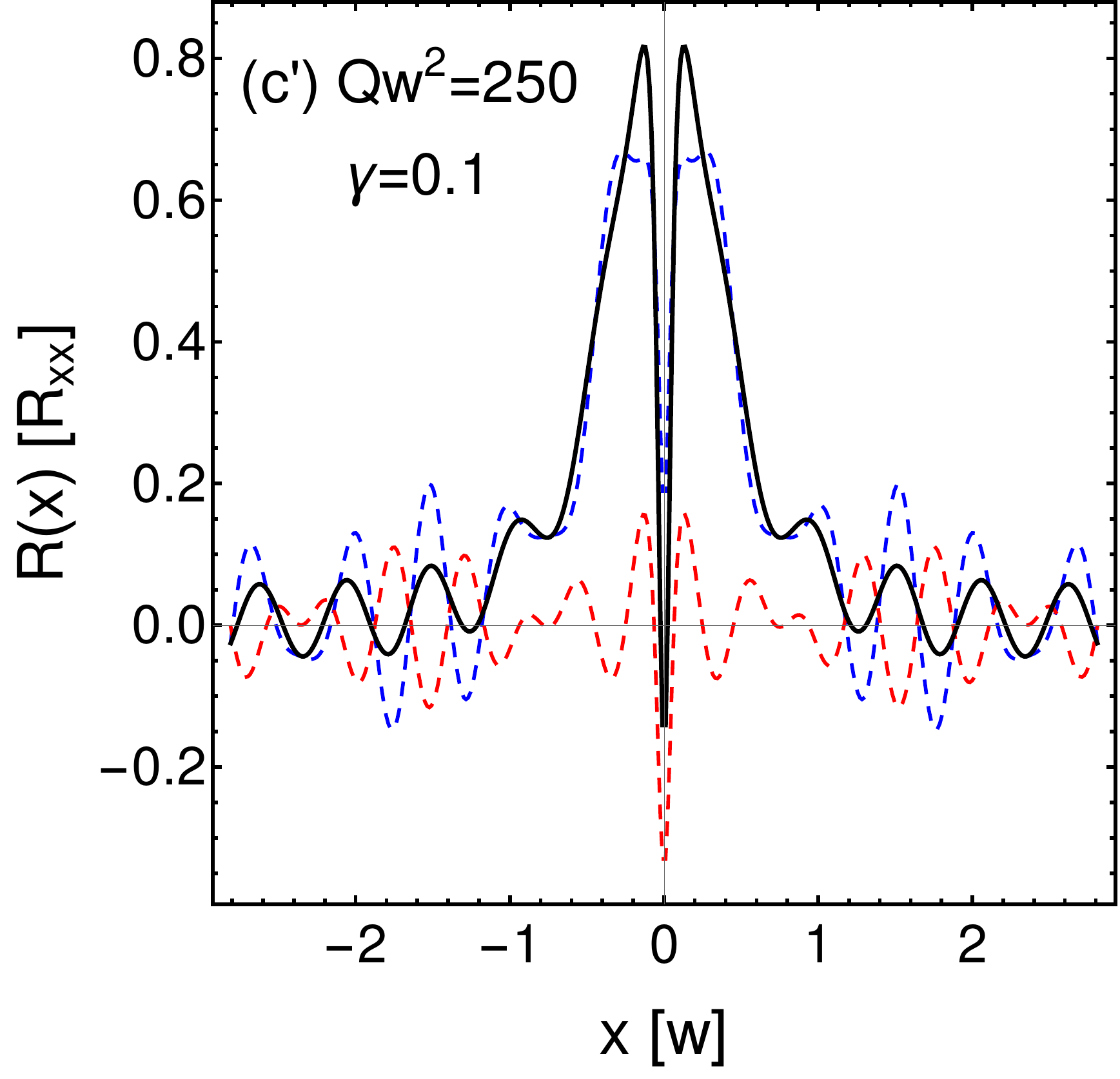}
\includegraphics[width=0.235\textwidth]{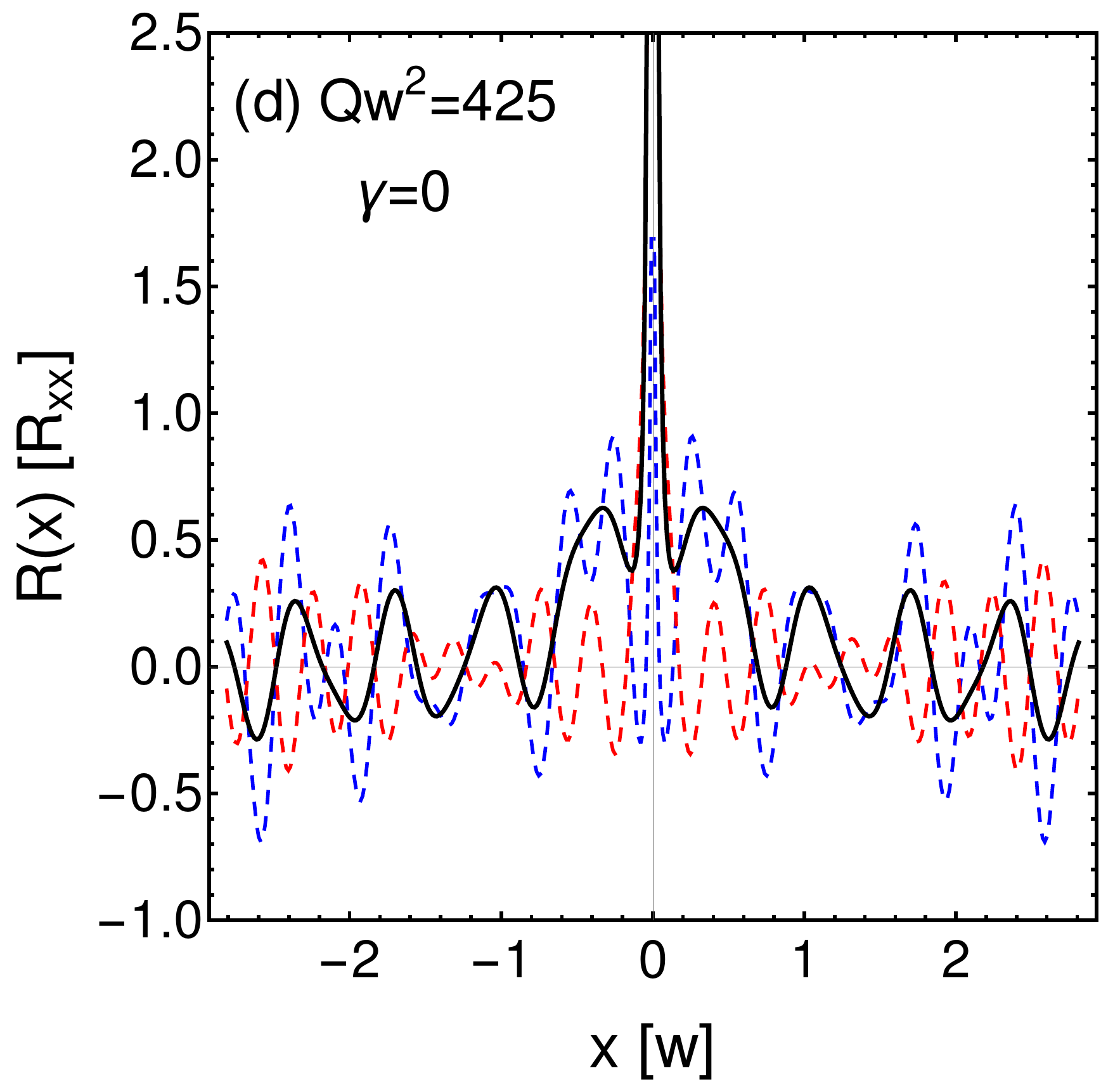} 
\includegraphics[width=0.235\textwidth]{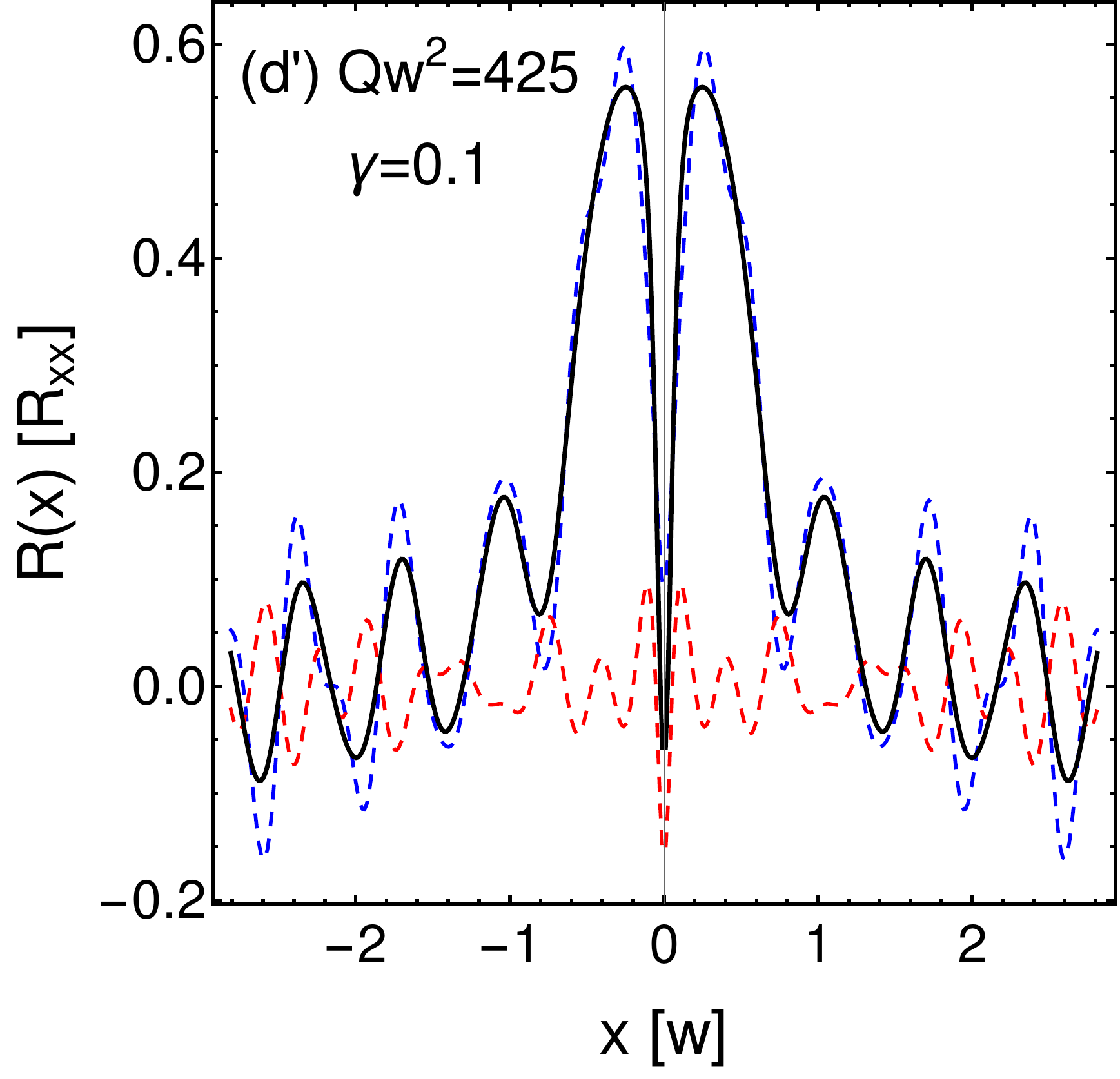}
\caption{Effective nonlocal resistivity $R(x)/R_{xx}$ in a strip of monolayer graphene corresponding to the current patterns in Fig.~\ref{fig2}. The blue and red dashed curves are bulk resistivity $R_0(x)$ and viscosity imbalance $R_{I}(x)$ contributions, respectively. We note that finite lead size strongly modifies the local resistivity $R(0)$ and, moreover, $R_0(0)>0$, $R_{I}(0)<0$, and $R(0) <0$. (a) and (a$^\prime$) for $Q < Q^\ast$, $R(x)$ exhibits a couple of sign reversals and decays rapidly as $x$ increases. (b)--(d) and (b$^\prime$)-(d$^\prime$) for $Q > Q^\ast$, $R(x)$ oscillates about zero as $x$ increases.}    
\label{fig3}
\end{figure*}

\subsection{Eddy pattern and nonlocal resistances} \label{Sec-Eddy}

We study the nonlocal response in the strip geometry defined by the area $0< y< w$ with transverse charge current $I$ injected and drained through a pair of contacts at $x=0$. We solve the streamfunction equation \eqref{stream-eq} analytically with the no-slip boundary condition $\d_y \varphi|_{y=0, w}=0$ and $-\d_x \varphi|_{y=0, w}=I \frac{\g}{\pi(\g^2+x^2)}$, where $\g>0$ describes the size of the contacts,
\be \label{phi-s}
\varphi(x,y) = -\frac{I}{\pi}  \int_{0}^{\infty} \! k^{-1} e^{-\g k} \sin{(kx)}  g(k,y;Q) dk,
\ee
where $g(k,y;Q)= [f(q,k,y)-f(k,q,y)]/[f(q,k,0)-f(k,q,0)]$ with $f(k,q,y)= k \sinh(kw/2) \cosh[q(y-1/2)]$ and $q=\sqrt{k^2-Q}$ for $k^2 \ge Q$ and $q=i\sqrt{Q-k^2}$ for $k^2 < Q$. We note that the dimensionless parameter $\mathsf{Q} = Qw^2$ fully determines the flow, and when $\mathsf{Q}  > \mathsf{Q}^\ast \approx 37.01$, $g(k,Q)$ processes simple poles, and the integral \eqref{phi-s} takes Cauchy principal values. The nonlocal resistivity is determined by the voltage across the width of the layer as a function of the $x$ coordinate, $R(x) \equiv [\del{V}(x,0)-\del{V}(x,w)]/I$, and does not depend on the Hall coefficients $R_{xy}$ and $C_{xy}$. We show the charge flows for various $\mathsf{Q} >0$ and lead sizes $\g$ in Fig.~\ref{fig2}. For $\mathsf{Q}  < \mathsf{Q}^\ast$, the amplitudes of vortices decay exponentially with distance to leads, $\ln|\mathbf{J}| \sim - \sqrt{Q^\ast-Q}|x|$, and are reduced for finite lead widths $\g>0$. For $\mathsf{Q} > \mathsf{Q}^\ast$, the vortices form eddy flows that are stable against finite lead width   
and possible different forms of boundary conditions (e.g., the no-stress boundary layer considered in Appendix \ref{App-Fourier}). Moreover, away from the leads  we observe that the number of vortices across the strip is odd, $2n-1$, where $n$ is the number of poles of $g(k)$ for $k \in [0,\sqrt{Q^\ast}]$. In Fig.~\ref{fig3} we show the nonlocal resistivity $R(x)$ corresponding to the current patterns in Fig.~\ref{fig2}. We note two important observations. (i) For $\mathsf{Q} < \mathsf{Q}^\ast$, $R(x)$ exhibits a couple of sign reversals and decays rapidly as $x$ increases, while for $\mathsf{Q} > \mathsf{Q}^\ast$, $R(x)$ oscillates about zero as $x$ increaseas. This coincides with the eddy picture. (ii) Finite lead size $\g >0$ modifies the local resistivity $R(x \sim 0)$, and one has $R_0(0)>0$, $R_{I}(0)<0$, and $R(0) <0$.

\section{Discussion and Outlook} \label{Sec-Fin}

A few comments are in order in relation to the results presented in this paper and in the context of recent related studies. 
The first point concerns terminology. In this work we adopted the concept of negative viscosity and used it in relation to the unconventional sign of the resistivity-to-viscosity ratio parameter $Q=R_{xx}/C_{xx}$ in Eq. \eqref{stream-eq} that determines the pattern of the macroscopic flow through the stream function. As is known from previously studied examples, the approximation based on the introduction of a single large-scale coordinate provides a successful description of the formation of regular eddy systems  (see, e.g., Fig. 4 from Ref. \cite{Sivashinsky-Turbul}). In the present case of an electron liquid in MLG, the underlying microscopic mechanism is completely different and comes from the coupling of charge modes to particle number and temperature imbalance modes. However, it is, in a sense, analogous to other historical findings where fluid flows couple to, e.g., magnetization modes \cite{Bacri} or some other modes in the system, which gives rise to the formation of stable vorticities via the effective negative viscosity effect (in particular, see Ref. \cite{Chechkin-Review} for a more detailed review of two-dimensional magnetohydrodynamic flows with negative viscosity). 

It has been shown in a recent work [\onlinecite{Alekseev-Counterflow}] that the interplay between viscosity and fast recombination in a two-component conductor (e.g., e-h plasma in MLG) leads to the appearance of current counterflows. In the geometry of the lateral transport current, the distribution of the edge currents in the transverse direction was found to possess a nontrivial spatial profile that consist of two stripelike regions: the outer stripe, which carries most of the current in the direction of the external electric field, and the inner stripe, with the counterflow. The functional form of the flow profile is a periodic function whose oscillatory part and decay part are controlled by the same scale (overdamped oscillation). We make the same observations concerning the importance of the interplay of viscous and relaxation effects but consider a different transport geometry and find a more pronounced regime of oscillatory vortex response. 

In addition, it was proposed earlier in Ref.  [\onlinecite{Mendoza}] that vorticities in the preturbulent regime could be observed in MLG provided there was a relatively high Reynolds number ($\sim 10$). The Strouhal number that measures the vortex shedding frequency was also estimated, with reasonable assumptions about the conditions of possible experiments. However, we discussed here a different kind of vortex response that crucially relies on carrier imbalance and occurs already at the level of linear hydrodynamics in a low Reynolds number regime (Poiseuille flow). We expect that the observability of eddies and related resistance oscillations should be accessible with existing high-quality hBN-MLG devices, but we realize that this could still be challenging. The reason is that the typical time-scale describing the generation/recombination processes is much longer than the e-e(h) equilibration time. Indeed, due to kinematic constraints, imbalance relaxation time requires multiparticle collisions. Close to neutrality the corresponding rate could be estimated as $\tau^{-1}_{\mathrm{imb}}\sim\alpha^4T$, up to some logarithmic factors $\ln(1/\alpha)$, which are clearly suppressed compared to the equilibration rate, $\tau^{-1}_{\mathrm{ee}}\sim\alpha^2T$, for the case of weak interaction $\alpha\ll1$.  However, at moderately strong interactions $\alpha\sim1$ the imbalance decay rate can be relatively high. In addition, it is strongly sensitive to electron--optical-phonon scattering (see Ref. [\onlinecite{Xie2016}] and estimates in Appendix \ref{App-Relaxation}). One should also keep in mind that the kinetic coefficient $\varsigma$, which in a way defines the imbalance-to-viscosity ratio $r$, is not solely governed by imbalance relaxation but is also strongly dependent on energy relaxation  processes via the inverse of $\lambda_{\alpha\beta}$ [Eq. \eqref{lambdas}], which mixes sectors of carrier imbalance and thermal modes and consequently favors higher values of $r$. 

Regarding the outlook, we wish to mention that the theory developed in this work may shed some light on the observed sign change of the Coulomb drag in a nonlocal measurement setup of graphene double layers [\onlinecite{Dean}], and the corresponding analysis will be presented in a separate work [\onlinecite{XFL}].             

\subsection*{Acknowledgments} 

We are grateful to M. Foster, K. C. Fong, D. Y. H. Ho, I. Gornyi, and B. Shklovskii for discussions. This work was financially supported by NSF Grant No. DMR-1606517, NSF CAREER Grant No. DMR-1653661, and a grant from the Wisconsin Alumni Research Foundation. 

\appendix

\section{Hydrodynamic theory}\label{App-Hydro}

We introduce the hydrodynamic equations of motion for massless Dirac fermions in relativistically covariant notation,
\begin{align}
& \d_a J^a = 0, \quad \d_a P^a = \mathcal{I}, \quad
\d_a \Theta^{ab} = - F^{ab} J_{b}/v_F + f^a, \label{rel-hyd}
\end{align}
where summation over repeated space-time indices $a,b \in \{0,1,2\}$ is assumed with $x^a \in \{v_F t,x,y \}$, $\d_a = \d/\d x^a$, and metric $g^{ab}=\mathrm{diag}\{+1,-1,-1\}$. 
In Eq.~\eqref{rel-hyd}, $J^a$ and $P^a$ denote the charge and quasiparticle three-current density, respectively, $\Theta^{ab}$ is the traceless energy-momentum tensor, and the Faraday tensor 
\be
F^{ab} = \begin{pmatrix} 
0   &    E_{x}        & E_{y} \\
-E_{x} &      0         &  v_F B \\ 
-E_{y} &  -v_F B    &  0
\end{pmatrix},
\ee
incorporating an in-plane electric fields $\mathbf{E}$ and a transverse magnetic field $\mathbf{B} = B \hat{\mathbf{z}}$. The quasiparticle imbalance flux $\mathcal{I}$ describes the e-h generation/recombination processes. The friction force density $f^a$ manifests the energy-momentum relaxations. Assuming fast intralayer equilibration of the carriers due to strong inelastic e-e and e-h Coulomb collisions, we express the current densities and energy-momentum tensor in terms of local thermodynamic variables; hydrodynamic three-velocity $U^a = \xi ( v_F,\mathbf{u} )$, with $\mathbf{u}$ being the fluid velocity and 
$\xi = (1-\mathbf{u}^2/v_F^2)^{-1/2}$ being the dilation factor; and dissipative derivations from local equilibrium:
\begin{align}
 J^a \equiv \rho U^a  + j^a, \quad  P^a \equiv n U^a  + p^a, \nonumber \\ 
\Theta^{ab}= h \left( U^a U^b/v_F^2 - g^{ab}/3 \right) + \theta^{ab}, \label{hyd-decomp}
\end{align}
where $\rho$, $n$, and $h$ are the proper (rest-frame) electric charge, carrier, and enthalpy density, respectively, and $j^a$, $p^a$, and $\theta^{ab}$ are the dissipative fluctuations of the charge current density, number current density, and energy-momentum tensor, respectively. Ensuring that the proper internal energy, electron and hole densities receive no dissipative corrections, the dissipative fluctuations are orthogonal to the fluid velocity: $U_{a} j^a =0$, $U_{a} p^a =0$, and $ U_{a} \theta^{ab} =0$. Applying the fluid velocity $U_{a}$ and the projector $\mathsf{P}_{ab} \equiv U_{a}  U_{b}/v_F^2-g_{ab}$ to the energy-momentum continuity equation in Eq.~\eqref{rel-hyd}, we obtain the energy evolution equation along $U_{a}$ and the momentum evolution equation perpendicular to it. Inserting the decomposition \eqref{hyd-decomp} and taking the  limit $\mathbf{u}^2/v_F^2 \ll 1$, we obtain the Navier-Stokes equation (1b) and entropy production equation (1c).

\section{Properties of the resistivity operator}\label{App-R}

In Eq.~\eqref{E-response} defining $\hat{K}_{\txa\txb} = \hat{\mathcal{K}}_{\txa \txb}(\mathsf{Z}=0)$ and decomposing $\hat{\mathcal{L}}_{\a\b} = \mathcal{L}_{0,\a\b} \hat{\mathbb{I}} + \mathcal{L}_{2,\a\b} \hat{\ep}$ and $\hat{K}_{\txa\txb} = K_{0,\txa\txb} \hat{\mathbb{I}} + K_{2,\txa\txb} \hat{\ep}$, we expand the resistivity operator in $\hat{D}$ and obtain
\begin{widetext}
\begin{align}
\hat{\mathcal{R}} \cdot \mathbf{J} = &\,\hat{\mathcal{L}}_{00} \cdot \mathbf{J}
- \left(\hat{\mathcal{L}}_{0 \txa} - \mathsf{Z}_{0 \txa} \hat{D}\right) \hat{K}_{\txa \txa_0} \left( \delta_{\txa_0 \txb} + \sum_{n=1}^{\infty} \mathsf{Z}_{\txa_0 \txb_1} \hat{D} \hat{K}_{\txb_1 \txa_1} \mathsf{Z}_{\txa_1 \txb_2} \hat{D} \hat{K}_{\txb_2 \txa_2} \cdots \mathsf{Z}_{\txa_{n-1} \txb_n} \hat{D} \hat{K}_{\txb_n \txb} \right) \hat{\mathcal{L}}_{\txb 0}\cdot \mathbf{J}, \nn \\
 = &\, \left( \hat{\mathcal{L}}_{00} -\hat{\mathcal{L}}_{0 \txa} \hat{K}_{\txa \txb} \hat{\mathcal{L}}_{\txb 0} \right) \cdot \mathbf{J} + \left( \mathsf{Z}_{0 \txa} \hat{\mathbb{I}} - \hat{\mathcal{L}}_{0 \txa_1} \hat{K}_{\txa_1 \txa_2} \mathsf{Z}_{\txa_2 \txa} \right)
\nabla^2 [(1 + \hat{K}_0 \hat{\mathsf{Z}} \nabla^2)^{-1}]_{\txa \txa_3}  \left( K_{0,\txa_3 \txb} \mathcal{L}_{2,\txb 0} +K_{2,\txa_3 \txb}\mathcal{L}_{0, \txb 0} \right) \hat{\ep} \cdot \mathbf{J}, 
\end{align}
\end{widetext}
where we have applied the identities due to the charge conservation, $\hat{D} \cdot \mathbf{J} =0$ and $\hat{D} \hat{\ep} \cdot  \mathbf{J} = \nabla^2 \hat{\ep} \cdot \mathbf{J}$.
(i) For $B=0$, $\hat{K}_{2} = \hat{\mathcal{L}}_{2} =0$ so that $\hat{\mathcal{R}}(B=0)=\mathbb{I}(\mathcal{L}_{0,00} -\mathcal{L}_{0,0 \txa} K_{0,\txa \txb} \mathcal{L}_{0,\txb 0})$.
(ii) At neutrality $\rho=0$, since $\mathsf{Z}_{0\mathrm{a}}=0$, $\Pi_{0 \mathrm{a},xx}=\hat{\Pi}_{\mathrm{a}0,xx}=0$, $\Pi_{00,xx}=\s_{00}^{-1}$, and $\hat{\Pi}_{\a \b,xy} = -B \left(\frac{T}{e h}\right)(\del_{\a 0} \del_{\b 2} + \del_{\a 2} \del_{\b 0})\hat{\ep}$, we have $\hat{\mathcal{R}} = \s_{00}^{-1} \hat{\mathbb{I}}- B^2 \left( \frac{T}{e h} \right)^2 \hat{\ep} \hat{\mathcal{K}}_{\mathrm{2}\mathrm{2}} \hat{\ep}$ and $\hat{\mathcal{R}} \cdot \mathbf{J} = \mathcal{R}_{xx} \mathbf{J}$ where $\mathcal{R}_{xx} = \s_{00}^{-1} + B^2 \left( \frac{T}{e h} \right)^2 \td{\mathcal{K}}_{22}$ with $[\hat{\td{\mathcal{K}}}^{-1}]_{\mathrm{a}\mathrm{b}} = \Pi_{\mathrm{a}\mathrm{b},xx} + \left( H_{\mathrm{a}\mathrm{b}} + \mathsf{Z}_{\mathrm{a}\mathrm{b}} \right)(-\nabla^2)$. 
(iii) The double-gradient operator can be decomposed as follows $\hat{D}(\l_1,\l_3) = (\nabla^2/2) \hat{\s}^0 +  \l_1 \d_x \d_y \hat{\s}^1 + \l_3 [(\d_x^2-\d_y^2)/2] \hat{\s}^3$, where $\hat{\s}^{1,2,3}$ are the $xy$-space Pauli matrices, $\hat{\s}^0 = \hat{\mathbb{I}}$, and $\l_{1,3}=1$ are the auxiliary parameters. We write the resistivity operator in the form
\be 
\hat{\mathcal{R}}(B,\l_1,\l_3) = \sum_{i =0}^3 \hat{\mathcal{R}}_i (B,\l_1,\l_3) \otimes \hat{\s}^i.  
\ee
Since the theory is invariant under the transforms $B \to -B$ and $x \leftrightarrow y$ or $B \to -B$ and $y \to -y$, which are represented by $\hat{\s}^1 \hat{\mathcal{R}}(-B,\l_1,-\l_3) \hat{\s}^1 = \hat{\mathcal{R}}(B,\l_1,\l_3)$ or $\hat{\s}^3 \hat{\mathcal{R}}(-B,-\l_1,\l_3) \hat{\s}^3 = \hat{\mathcal{R}}(B,\l_1,\l_3)$, we readily have the symmetries 
\begin{align} 
&\hat{\mathcal{R}}_{0,1} (-B,  \l_1, -\l_3) = \hat{\mathcal{R}}_{0,1} (B,\l_1,\l_3), \nonumber \\ 
&\hat{\mathcal{R}}_{2,3} (-B, \l_1, -\l_3) = -\hat{\mathcal{R}}_{2,3} (B,\l_1,\l_3),  \nonumber \\
&\hat{\mathcal{R}}_{0,3} (-B, -\l_1,  \l_3) = \hat{\mathcal{R}}_{0,3} (B,\l_1,\l_3), \nonumber \\ 
&\hat{\mathcal{R}}_{1,2} (-B, -\l_1, \l_3) = -\hat{\mathcal{R}}_{1,2} (B,\l_1,\l_3).
\end{align}
Furthermore, for $\l_1=0$ and $\l_3=0$, we respectively obtain higher symmetries,
\begin{align} \label{l1-0}
&\hat{\mathcal{R}}_{0,3}(-B,0,\l_3)=\hat{\mathcal{R}}_{0,3}(B,0,\l_3), \nonumber \\ 
&\hat{\mathcal{R}}_{1,2}(-B,0,\l_3)=-\hat{\mathcal{R}}_{1,2}(B,0,\l_3), \nonumber \\
&\hat{\mathcal{R}}_{0,1}(-B,\l_1,0)=\hat{\mathcal{R}}_{0,1}(B,\l_1,0), \nonumber \\ 
&\hat{\mathcal{R}}_{2,3}(-B,\l_1,0)=-\hat{\mathcal{R}}_{2,3}(B,\l_1,0),
\end{align}  
The equations in \eqref{l1-0} imply that in the absence of the charge conservation the longitudinal (Hall) components could involve odd (even) powers of $B$ induced by the imbalance effects (due to the transverse response $D_{xy}=\d_x\d_y$ ).

\section{Derivation of the stream function equation}\label{App-Expansion}

For weak inhomogeneity we expand the resistivity operator up to the first order in the differential operators $\{\nabla^2, \hat{D}\}$, $\hat{\mathcal{R}} = \hat{R} + \del{\hat{R}}+ \mathcal{O}(\nabla^4, \hat{D}^2)$, where $\hat{R}$ are the resistivity matrices for infinitely large homogeneous systems,
\be
\hat{R} = \hat{\Pi}_{00}-\hat{\Pi}_{0 \mathrm{a}} \hat{\mathsf{K}}_{\mathrm{a} \mathrm{b}} \hat{\Pi}_{\mathrm{b} 0} = R_{xx} \hat{\mathbb{I}} + R_{xy} \hat{\ep},
\ee
with $[\hat{\mathsf{K}}^{-1}]_{\mathrm{a} \mathrm{b}} = \hat{\Pi}_{\mathrm{a} \mathrm{b}}$. The inhomogeneity corrections have two parts, $\del{\hat{R}} = \del \hat{R}_\l+\del \hat{R}_\eta$. 
Defining $\hat{X}_{a 0} \equiv \hat{\mathsf{K}}_{\mathrm{a} \mathrm{b}} \hat{\Pi}_{\mathrm{b} 0}$ and its transpose $\hat{X}^T_{0 \mathrm{a}} = \hat{\Pi}_{0 \mathrm{b}} \hat{\mathsf{K}}_{\mathrm{b} \mathrm{a}}$, we obtain the imbalance and bulk viscosity corrections and shear viscosity corrections
\begin{align}
& \del \hat{R}_\eta = -\big( H_{00} \hat{\mathbb{I}} - H_{0 \mathrm{a}} \hat{X}_{\txa 0} - \hat{X}^T_{0 \mathrm{a}} H_{\mathrm{a}0}   +   \hat{X}^T_{0 \mathrm{a}} H_{\mathrm{a} \mathrm{b}} \hat{X}_{\mathrm{b} 0}\big) \nabla^2\nonumber 
\\ &= ( C_{0} \hat{\mathbb{I}} + C_{2} \hat{\ep} ) \nabla^2, \label{eta-cort} \\ 
& \del \hat{R}_{\l} = \big( \mathsf{Z}_{0 \mathrm{b}} \hat{\mathbb{I}}- \hat{X}^T_{0 \mathrm{a}} \mathsf{Z}_{\mathrm{a} \mathrm{b}} \big) X_{2,\mathrm{b} 0} \nabla^2 = ( C_{1} \hat{\mathbb{I}} + C_{3} \hat{\ep} ) \nabla^2, \label{lam-cort} 
\end{align}
where we have assumed $\hat{X}_{\mathrm{a} 0} = X_{0,\mathrm{a} 0} \hat{\mathbb{I}} +X_{2,\mathrm{a} 0} \hat{\ep}$. Here $\big\{ R_{xx}, \, C_{0}, \, C_{1} \big\}$ and $\big\{ R_{xy}, \, C_{2}, \, C_{3}\big\}$ are even and odd functions of $B$, respectively. Finally, we have
\begin{align}
&\mathcal{R}_{xx}= \mathcal{R}_{yy}= R_{xx} + C_{xx} \nabla^2, \nonumber \\ 
&\mathcal{R}_{xy} =-\mathcal{R}_{yx}= R_{xy}  + C_{xy} \nabla^2,
\end{align}
where $C_{xx}= C_{0} + C_{1}$ and $C_{xy}= C_{2} + C_{3}$. The stream function equation the reads as Eq.~\eqref{stream-eq} in the main text.

\section{Carrier and energy relaxation coefficients}\label{App-Relaxation}

We evaluate the relaxation coefficients $\hat{\l}$ and $\tau_\mathrm{el}$ in Eq.~(4) using kinetic theory, and the definition of the collision integrals can be found in Ref. [\onlinecite{Xie2016}]. The carrier-population imbalance relaxation $\l_{11}$ are caused by both optical phonon scattering and three-body Coulomb collisions $\l_{11} = \l_{11}^{\mathrm{ph}} + \l_{11}^{c}$, and $\l_{12,21,22}$ are caused by only optical phonons. The optical phonon scattering leads to
the carrier imbalance and energy relaxation coefficients,
\begin{align}
&\l_{11}^{\mathrm{ph}}= \frac{8 e^2 v_F^2 \a_\mathrm{ph}}{T} \mathrm{csch}\left( \frac{\w_{A'}}{2T}\right)\int \frac{d^2 \bp d^2 \bq}{(2\pi)^4} \left( \frac{1 - \hat{\bp}\cdot\hat{\bq}}{2} \right) \nn \\ 
&\delta( \ep_\bp+\ep_\bq - \w_{A'}) O_{p,q}^{+1,-1}, \quad \l_{12,21}  = \frac{\w_{A'}}{2T} \l_{11}^{\mathrm{ph}}, \nn \\
&\l_{22} = \left( \frac{\w_{A'}}{2T} \right)^2 \l_{11}^{\mathrm{ph}} +  \frac{2 e^2 \w_{A'}^2 v_F^2 \a_\mathrm{ph}}{T^2} \mathrm{csch}\left( \frac{\w_{A'}}{2T}\right) \nn \\  
&\int \frac{d^2 \bp d^2 \bq}{(2\pi)^4}\left( \frac{1 - \hat{\bp}\cdot\hat{\bq}}{2} \right) \delta( \ep_\bp-\ep_\bq - \w_{A'})  \sum_{s}O_{p,q}^{s,s}.
\end{align}
Here $O_{p_1,p_2}^{s_1,s_2} =(1/4)\mathrm{sech}{[\b(\ep_{\bp_1} - s_1 \mu)/2]} \mathrm{sech}{[\b(\ep_{\bp_2} -s_2 \mu) /2]}$. The dimensionless effective electron-phonon scattering strength is $\a_\mathrm{ph}=(2\pi)^2 \b_{A^\prime}^2 s_0/v_F^2 M \w_{A^\prime}$, with $M = 2.0 \times 10^{-23} \, \mathrm{g}$ being the carbon atom mass, $s_0 = 2.62 \, \AA^2$ being the area per carbon atom, $\omega_{A^\prime}$ being the optical phonon frequency, and $\beta_{A^\prime}$ being the electron-phonon coupling. We estimate the three-body collision contribution $\l_{11}^c \approx 4 \ln(2) \a^4 e^2 T^2/\pi v_F^2$[\onlinecite{Lucas2018}]. The momentum relaxation scattering rate reads $\tau_\mathrm{el}^{-1} = \tau_\mathrm{imp}^{-1} + \tau_\mathrm{ph}^{-1}$ where the individual rates due to impurities and optical-phonon scatterings are 
\begin{align} 
&\tau_\mathrm{imp}^{-1}  = \frac{2 \pi v_F^2}{T h}  \int\frac{d^2 \bp d^2 \bq}{(2\pi)^4} \delta( \ep_\bp-\ep_\bq) |\bp-\bq|^2 V_\mathrm{imp}(\bp,\bq) \sum_{s}O_{p,p}^{s,s}, \nn \\
&\tau_\mathrm{ph}^{-1} =  \frac{v_F^4 \a_\mathrm{ph}}{T h} \mathrm{csch}\left( \frac{\w_{A'}}{2T}\right) \int\frac{d^2 \bp d^2 \bq}{(2\pi)^4} \left( \frac{1 - \hat{\bp}\cdot\hat{\bq}}{2} \right) \nn \\ 
&\sum_{s,s'} \delta( s \ep_\bp - s \w_{A'} - s' \ep_\bq) |s \bp- s' \bq|^2 O_{p,q}^{s,s'}, 
\end{align}
where $V_\mathrm{imp}(\bp,\bq) = |\rho_\mathrm{min}/e|(1 + \hat{\bp}\cdot\hat{\bq}) | U_\mathrm{eff}(0,|\bp-\bq|)|^2/2$ describes the Coulomb impurity scattering strength, with $|\rho_\mathrm{min}/e|$ being the charged impurity concentration and $U_\mathrm{eff}(\w,q)$ being the random-phase-approximation screened Coulomb potential. In calculation we take $\beta_{A^\prime} = 10 \, \mathrm{eV}/\AA$ and $\w_{A^\prime} = 1740 \, \mathrm{K}$, so that $\a_\mathrm{ph} \approx 2.2$, and $|\rho_\mathrm{min}/e| =5 \times 10^{9} \mathrm{cm}^{-2}$.

\begin{figure*}
\includegraphics[width=0.4\textwidth]{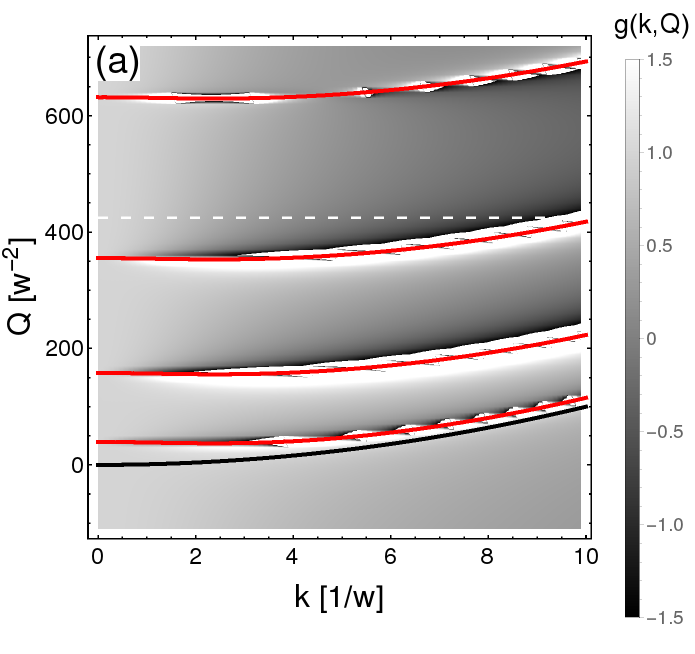} 
\includegraphics[width=0.36\textwidth]{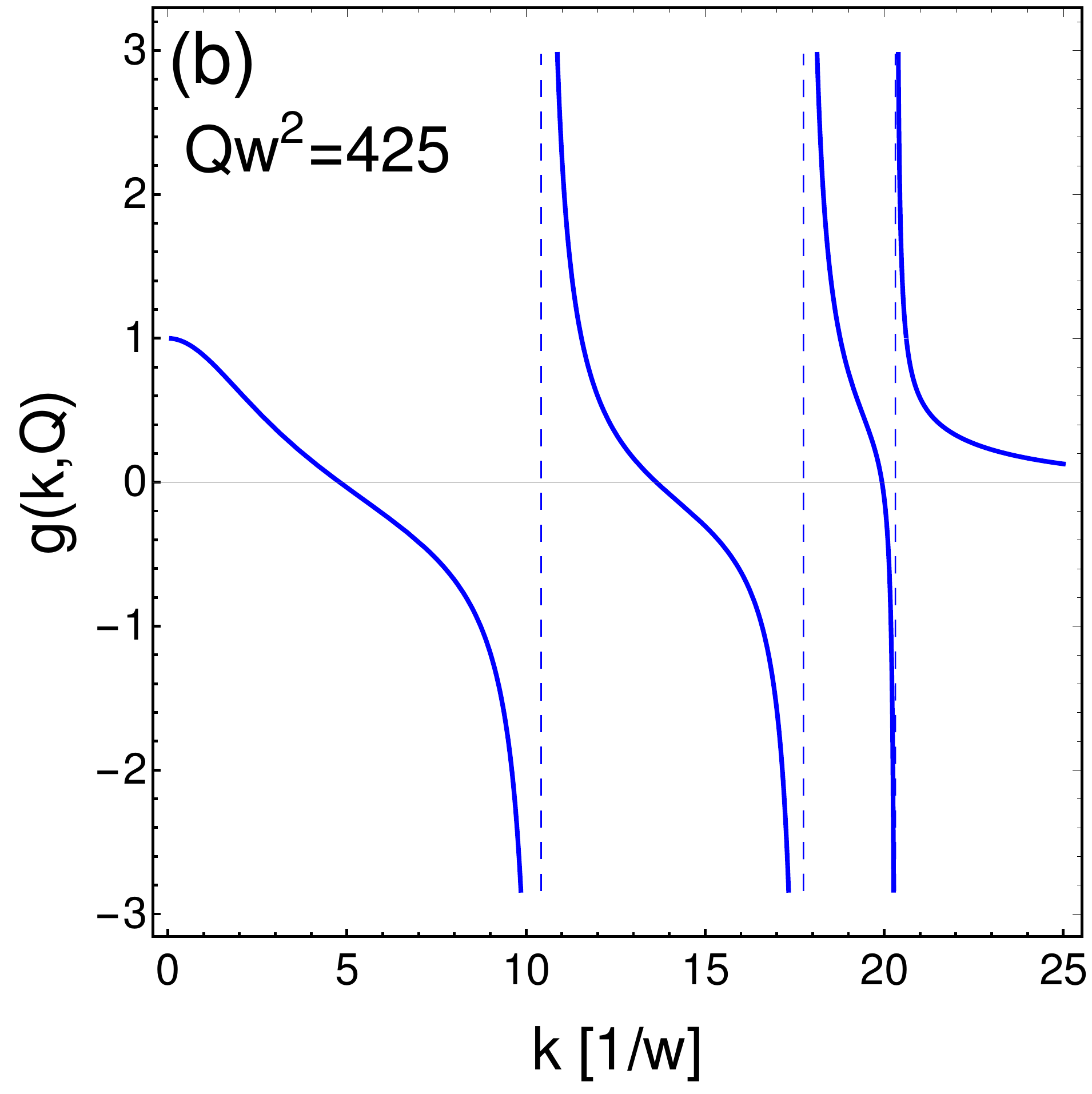}
\caption{
Schematic plots for the analytical properties of the function $g(k,Q)$ in Eq.~\eqref{g-f}. We take $w=1$ and $y=0.2$. (a) The black curve indicates $Q=k^2$, and the red curves show the singularities of $g(k,Q)$. (b) $g(k)$ for $Q=425$ [along the white dashed line in panel (a)]. The singularities are simple poles.}    
 \label{fig-S1}
\end{figure*} 

\section{Solution of the stream function equation in a strip geometry}\label{App-Fourier}

\subsection{No-slip boundary condition}

After the partial Fourier transform in the $x$ direction $\varphi(x,y) = \int\!\frac{dk}{2\pi}\, \varphi_{k}(y)  e^{i k x}$, Eq.~\eqref{stream-eq} becomes
\begin{align} \label{PFT}
&(\d_y^2-k^2) (\d_y^2-q^2) \varphi_{k}(y)=0, \nn \\ 
&\varphi_{k}|_{y=0,w}=i I(k)/k, \quad \d_y\varphi_{k}|_{y=0,w}=0,  
\end{align}
where $q = \sqrt{k^2-Q}$ for $k^2 \ge Q$ and $q = i \sqrt{Q-k^2}$ for $k^2<Q$ and $I_{k} = e^{-\g k}$. The general solution takes the form
\be 
\varphi_{k} = \frac{i I(k)}{k} \sum_{s=\pm 1} (a_{s} e^{s k y }+ b_{s} e^{s q y }).
\ee
Matching the boundary conditions, we determine the coefficients by

\begin{widetext}
\begin{align}
\sum_{s=\pm 1} (a_{s} + b_{s}) =1, \quad  \sum_{s=\pm 1} (a_{s} e^{s k w }+ b_{s} e^{s q w }) =1, \quad
\sum_{s=\pm 1} (s k a_{s} + s q b_{s})= \sum_{s=\pm 1} (s k a_{s} e^{s k w } + s q b_{s} e^{s q w }) = 0. 
\end{align}
Solving $\{ a_{s},b_{s} \}$ we obtain 
\begin{align}
a_{j,+} = \frac{(e^{q w}-1)q}{M(k, q)}, \quad
a_{j,-} = \frac{e^{kw}(e^{q w}-1)q}{M(k, q)},\quad
b_{j,+} = \frac{(1-e^{k w})k}{M(k, q)}, \quad
b_{j,-} = \frac{e^{q w}(1-e^{k w})k}{M(k, q)},
\end{align} 
where $M(k, q) = (k-q)[1-e^{(k+q)w}] + (k+q)(e^{q w} - e^{k w})$. The Fourier transform gives
\begin{align}
& \varphi(x,y) = - \int_{-\infty}^{\infty}\! \frac{dk I(k) e^{i k x}}{2 \pi i k } g(k,Q) = -\frac{1}{\pi} \int_{0}^{\infty}\! \frac{dk I(k) \sin{(kx)}}{k} g(k,Q_j), \label{phi-def} \\
& g(k,Q) =
\frac{q \sinh \left(\frac{q w}{2}\right) \cosh \left[k \left(y-\frac{w}{2}\right)\right]-k \sinh \left(\frac{k w}{2}\right) \cosh \left[q
   \left(y-\frac{w}{2}\right)\right]}{q \cosh \left(\frac{k w}{2}\right) \sinh \left(\frac{q w}{2}\right)-k \sinh \left(\frac{k w}{2}\right) \cosh
   \left(\frac{q w}{2}\right)}. \label{g-f}
\end{align}

In Fig.~\ref{fig-S1} we show the analytic properties of the integrand $g(k,Q)$ where we set $w=1$. We find that, when $Q > Q^\ast \approx 37.01$, $g(k,Q)$ has simple poles and the integral \eqref{phi-def} takes Cauchy principal values. 
The voltage drops between $(x,0)$ and $(x,w)$, $\D V_i(x) \equiv \del V(x,0)-\del V(x,w)$, read
\begin{align} 
\D V(x) = &\, \int_{0}^{w} dy \mathcal{E}_{y} = \int_{0}^{w} dy \left( R_{xx} + C_{xx} \nabla^2   \right) J_{y} \nn \\
    = &\, - \frac{2}{\pi} \int_{0}^{\infty}\!dk \left(  R_{xx} - C_{xx} k^2 \right) I(k) \cos{(kx)} \frac{ Q \sinh\left( \frac{k w}{2} \right)\sinh\left( \frac{q w}{2} \right)/(k q)}{q \cosh \left(\frac{k w}{2}\right) \sinh \left(\frac{q w}{2}\right)-k \sinh \left(\frac{k w}{2}\right) \cosh\left(\frac{q w}{2}\right)}.
\end{align}
The nonlocal resistance is defined by $R(x) = \D V(x)/ I$. For point-like leads $\g=0$, in the Stokes and Ohm limit, we obtain
\be
R(x) = \begin{cases} &  \frac{C_{xx}}{w^2} f(x/w), \quad f(z) = -\frac{8}{\pi} \int_{0}^{\infty}\!d k \frac{k \cos{(kz)} \sinh^2(k/2)}{k+\sinh{k}}, \quad R_{xx} = 0, \\
&  \frac{2}{\pi} R_{xx} \int_{0}^\infty \! \frac{dk \cos{(kx)}}{k} \tanh\left( \frac{k w}{2} \right) =  R_{xx} \frac{2}{\pi} \ln\left| \coth\left( \frac{\pi x}{2w}\right) \right|, \quad C_{xx} = 0.
\end{cases}  
\ee

\subsection{No-stress boundary conditions} 

For the no-stress boundary conditions $\d_y^2 \varphi|_{y=0,w}=0$ and $-\d_x \varphi|_{y=0,w}= I(x) $, we obtain 
\begin{align}
& \sum_{s=\pm 1} (a_{s} + b_{s})=1, \quad  \sum_{s=\pm 1} (a_{s} e^{s k w }+ b_{s} e^{s q w }) =1, \nn \\
& \sum_{s=\pm 1} ( a_{s} k^2 + b_{s} q^2) = \sum_{s=\pm 1} (a_{s} k^2 e^{s k w } +  b_{s} q^2 e^{s q w }) = 0. 
\end{align}
Solving $\{ a_{s},b_{s} \}$, we obtain 
\begin{align}
& a_{+} = -\frac{k^2-Q}{Q}\frac{1}{1+e^{kw}}, \quad
a_{-} = -\frac{k^2-Q}{Q}\frac{1}{1+e^{-kw}},\quad 
b_{+} = \frac{k^2}{Q}\frac{1}{1+e^{q w}}, \quad
b_{-} = \frac{k^2}{Q}\frac{1}{1+e^{-q w}}, \nn \\
& \varphi(x,y) = -\frac{1}{\pi} \int_{0}^{\infty}\! \frac{dk I(k) \sin{(kx)}}{k} g'(k,Q), \quad g'(k,Q) = \frac{1}{Q^2} \left\{ 
k^2 \frac{\cosh[q(y-1/2)]}{\cosh(q w/2)} - q^2 \frac{\cosh[k(y-1/2)]}{\cosh(kw/2)} \right\}. \label{g2-f}
\end{align}
The analytical properties of $g'(k,Q)$ [Eq.~\eqref{g2-f}] are qualitatively identical to those of $g(k,Q)$ [Eq.~\ref{g-f}]: For $Q>\pi^2$, $g'(k,Q)$ has simple poles at $k=k_n^\ast$, where $k_n^\ast = \sqrt{Q-[(2n+1)\pi]^2}$ for $ 0 \le n \le \lfloor (Q/\pi - 1)/2 \rfloor$ ($w=1$). Hence, the eddy flow pattern is presumably robust against boundary conductions. 
\end{widetext}

\end{document}